\documentclass[aps,pre,amsmath,singlecolumn,showpacs,11pt]{revtex4-1}
\usepackage{amsmath}
\usepackage{array}
\usepackage{amsfonts}
\usepackage{graphicx}
\usepackage{hyperref}
\usepackage{enumerate}
\usepackage{mathtools}
\usepackage{verbatim}
\usepackage{epstopdf}
\usepackage{subfigure}
\usepackage{latexsym}
\usepackage{dcolumn}
\usepackage{comment}
\usepackage{epsf}
\usepackage{float}
\DeclareMathSymbol{\mlq}{\mathord}{operators}{``}
\DeclareMathSymbol{\mrq}{\mathord}{operators}{`'}

\usepackage{color}

\usepackage{xcolor} 

\begin{document}
\title{Accuracy of a one-dimensional reduction of dynamical systems on networks}
\author{Prosenjit Kundu$^1$} 
\author{Hiroshi Kori$^2$}
\author{Naoki Masuda$^{1,3}$}
\email{naokimas@gmail.com}

\affiliation{$^1$Department of Mathematics, State University of New York at Buffalo, New York 14260-2900, USA }
\affiliation{$^2$Department of Complexity Science and Engineering, The University of Tokyo, Chiba 277-8561, Japan}
\affiliation{$^3$Computational and Data-Enabled Science and Engineering Program, State University of New York at Buffalo, Buffalo, New York 14260-5030, USA}
\date{\today}

\begin{abstract}
Resilience is an ability of a system with which the system can adjust its activity to maintain its functionality when it is perturbed. To study resilience of dynamics on networks, Gao {\it et al.} [Nature, {\bf{530}}, 307 (2016)] proposed a theoretical framework to  reduce dynamical systems on networks, which are high dimensional in general, to one-dimensional dynamical systems.  The accuracy of this one-dimensional reduction relies on several assumption in addition to the assumption that the network has a  negligible degree correlation. In the present study, we analyze the accuracy of the one-dimensional reduction assuming networks without degree correlation. We do so mainly through examining the validity of the individual assumptions underlying the method. Across five dynamical system models, we find that
the accuracy of the one-dimensional reduction hinges on the spread of the equilibrium value of the state variable across the nodes  in most cases. Specifically, the one-dimensional reduction tends to be accurate when the dispersion of the node's state is small. We also find that the correlation between the node's state and the node's degree, which is common for various dynamical systems on networks, is unrelated to the accuracy of the one-dimensional reduction.
\end{abstract}

 \maketitle
\section{Introduction}
A definition of resilience is the property of the system to be able to adjust its activity to retain its functionality  when some error or failure occurs \cite{Hosseini_RESS2016, Scheffer_book2009, Liu_Arxiv2020, Zhang_PRE2020}. Ecology \cite{Ungar_ES2018, Folke_EcoSoc2016}, human physiology \cite{Venegas_Nature2005}, cell biology \cite{Karlebach_NRMCB2008}, food security \cite{Barthel_EcoEcon2013, Suweis_PNAS2015}, finance \cite{Schweitzer_Science2009,Gualdi_JEDC2015}, and psychopathology \cite{Masten_ARCP2021,Wichers_PsyMed2019} are some of the areas where the resilience of systems has been studied.  A loss of resilience may give rise to a catastrophic outcome or breakdown, such as mass extinction in ecological systems \cite{Allesina_Nature2012, Suweis_NatComm2015, Grilli_NatComm2017}, blackouts in power grids \cite{Zhu_IEEE2014, Simpson_NatComm2016}, outbreaks of infectious diseases \cite{Drake_PlosCompBio2019, Massaro_SciRep2018}, crashes in financial markets \cite{Schweitzer_Science2009,Gualdi_JEDC2015}, and mental disorders \cite{Lunansky_EJP2020, Masten_ARCP2021}. The occurrence of such critical regime shifts bears some universality but depends on the type of dynamics and perturbations applied.  
Quantitatively anticipation of critical regime shifts in complex dynamical systems has been actively studied \cite{Scheffer_Nature2009, Scheffer_Science2012, Boettiger_Nature2013, Dablander_PsyArXiv2020, Liu_Arxiv2020}.  

Nonlinear dynamical systems on networks provide a useful language with which to investigate resilience in complex systems. In that framework, one assumes that dynamical elements occupy nodes and interact with adjacent nodes. If the state of each node is a one-dimensional dynamical variable, we typically examine the behavior of $N$ coupled differential equations, where $N$ is the number of nodes in the network. Alternatively, we may be interested in the case in which each node is a higher-dimensional dynamical system. In either case, one asks whether perturbations applied to the states of some nodes or removal of some edges, for example, cause catastrophic outcomes, i.e., drastic changes in the equilibrium of the entire $N$-dimensional dynamical system. 
For an ecosystem, a drastic change is typically from a stable equilibrium
corresponding to the coexistence of various species to a mass extinction of several species \cite{Morone_Nat_Phys2019,  Ceballos_PNAS2017}.
%
%
Crucially, the size of the perturbation that a dynamical system on a network can tolerate, which is an operational definition of the network resilience \cite{Liu_Arxiv2020}, depends on the structure of the network. For example, a change in the network structure induced by the removal of a fraction of nodes and the associated edges may decrease the resilience of the system  \cite{Gao_Energies2015, Gao_Nature2016, Fan_NJP2018}.

Understanding and predicting the resilience of dynamical systems on networks is a challenging task because of its high dimensionality and possible complexity in the network structure.
One principled approach to the study of network resilience is to reduce the $N$-dimensional dynamical system, assuming that each of the $N$ nodes has just one dynamical variable, to a tractable low-dimensional dynamical system. Gao {\it et al.} presented a theory to reduce any $N$-dimensional dynamical system belonging to a certain class to a one-dimensional dynamical system that approximates the dynamics of a linear weighted average of the $N$ state variables each of which is associated with a node \cite{Gao_Nature2016}. We refer to this method as the GBB (Gao-Barzel-Barab{\'a}si) reduction in the following text. The derived one-dimensional dynamical system is analytically tractable and approximates the behavior of the original $N$-dimensional dynamical system in terms of, for example, the position of the stable equilibria. The GBB  reduction is reasonably accurate when the number of nodes is larger than a critical value 
 \cite{Tu_PRE2017}. The reduction can also be accurate when the interaction matrix (i.e., the matrix whose entries are the edge weight) is a mixture of positive and negative weights  \cite{Tu_PRE2017}. In fact, the GBB  reduction assumes uncorrelated unipartite networks, whereas many ecosystems representing mutualistic interaction correspond to bipartite interaction networks. For bipartite networks, reduction to a two-dimensional dynamical system is a more accurate approach \cite{Jiang_PNAS2018,Laurence_PRX2019,Jiang_JRSI2019}. 
The GBB  reduction has also been extended to
%
%
the case in which the functional form of the individual node's dynamics and that of the influence of nodes on their adjacent nodes depend on the node \cite{Tu_IScience2021}. In another explicit work Thibeault {\it et al.} proposed a generalized technique where the high-dimensional dynamical systems are reduced to a low-dimensional dynamical systems by constructing a reduction matrix \cite{Thibeault_PRR2020}. This reduction technique is also applicable to the dynamical systems with diffusive kind of coupling functions as well  \cite{Thibeault_PRR2020}.
%

Despite these and other developments, the conditions under which these reduction methods are accurate at describing the original high-dimensional dynamics on networks are not sufficiently clear. By restricting our focus to the GBB reduction, we investigate its applicability in quantitative terms in the present study. In particular, we decompose the source of approximation error into different factors and investigate their contributions to the error for five types of dynamical systems. For each of the five considered dynamical system we analyze the approximations used for this reduction is shown in the next section. For all the dynamical systems we find out which property of the dynamical system leads to a accurate reduction.

\section{Dynamical systems on networks and their one-dimensional reduction}

We consider dynamical systems on networks considered in Ref.~\cite{Gao_Nature2016}, which are of the form
\begin{eqnarray}
\frac{dx_i}{dt}=F(x_i) +\sum_{j=1}^N A_{ij}G(x_i,x_j),
\label{multi_eq1}
\end{eqnarray}
where $x_i \in \mathbb{R}$ represents the dynamical state of the $i$th node, $N$ is the number of nodes,  $F(x_i)$ represents the intrinsic dynamics of the node,  $G(x_i,x_j)$ represents the influence of the $j$th node on the $i$th node, and  $A_{ij}$ is the weight of the directed edge $(j,i)$. 

Here we briefly describe the GBB  reduction of the $N$-dimensional dynamics described by Eq.~\eqref{multi_eq1} \cite{Gao_Nature2016}.
We first consider an arbitrary  scalar quantity $y_i$ associated with the $i$th node. To analyze Eq.~\eqref{multi_eq1}, we set $y_j(x_i)=G(x_i,x_j)$ and write
\begin{eqnarray}
\sum_{j=1}^{N} A_{ij}G(x_i,x_j)= s_i^{\mathrm{in}}\langle y_j(x_i)\rangle_{j~\mathrm{nn~of}~i},
\label{ynn1}
\end{eqnarray}
where $s_i^{\mathrm{in}}=\sum_{j=1}^N A_{ij}$ is the weighted in-degree (also called the in-strength) of the $i$th node, and $\langle y_j(x_i)\rangle_{j~\mathrm{nn~of}~i}= \sum_{j=1}^{N} A_{ij}G(x_i,x_j)/s_i^{\mathrm{in}}$ is the weighted average of $G(x_i, x_j)$ over the in-neighbors of the $i$th node, where the weight for the averaging is given by $A_{ij}$.  
We use the heterogeneous meanfield approximation  for $\langle y_j(x_i)\rangle _{j~\mathrm{nn~of}~i}$ as follows. Regardless of node $i$, we assume that its in-neighbor is $j$ with a probability proportional to $j$'s out-degree. Then, the corresponding weighted average of $y_j(x_i)$ for any $i$ is given by
\begin{eqnarray}
\langle y_j(x_i)\rangle_{\mathrm{HM}}=\frac{\frac{1}{N}\sum_{j=1}^N s_j^{\mathrm{out}}y_j(x_i)}{\frac{1}{N}\sum_{j=1}^N s_j^{\mathrm{out}}},
\label{ynn}
\end{eqnarray}
where $s_j^{\mathrm{out}}=\sum_{i=1}^N A_{ij}$ is the weighted  out-degree (also called the out-strength) of node $j$.
If the degree correlation of the network given by $A$ is small, then the neighborhood of all the nodes is considered to be statistically identical. 
In this case, one can write 
\begin{eqnarray}
\langle y_j(x_i)\rangle_{j~\mathrm{nn~of}~i}\approx \langle y_j(x_i)\rangle_{\mathrm{HM}},
\label{ynn2}
\end{eqnarray}
for all $i$, where $\approx$ represents ``approximately equal to".
Now, to formalize the approximation, we define the operator $\mathcal{L}$ by 
\begin{eqnarray}
\mathcal{L}(\mathbf{y})=\frac{\mathbf{1}^{\top}A\mathbf{y}}{\mathbf{1}^{\top}A\mathbf{1}}=\frac{\sum_{i=1}^N\sum_{j=1}^NA_{ij}y_j}{\sum_{i=1}^N\sum_{j=1}^NA_{ij}}&=&\frac{\frac{1}{N}\sum_{j=1}^N s_j^{\mathrm{out}}y_j}{\frac{1}{N}\sum_{j=1}^N s_j^{\mathrm{out}}}\nonumber \\ &=&\frac{\langle {s}^{\mathrm{out}}{y}\rangle}{\langle {s}^{\mathrm{out}} \rangle},
\label{ynn3}
\end{eqnarray}
where  $\mathbf{y} = (y_1, \ldots, y_N)^{\top}$, $\mathbf{1}=(1,\dots,1)^{\top}$, ${}^{\top}$ represents the transposition, and $\langle \cdot \rangle$ without a subscript represents the unweighted average over all nodes.
Using Eqs.~\eqref{ynn1}, \eqref{ynn}, \eqref{ynn2}, and \eqref{ynn3}, one can write Eq.~\eqref{multi_eq1} as
\begin{eqnarray}
\frac{dx_i}{dt} \approx F(x_i) + s_i^{\mathrm{in}}\mathcal{L}(G(x_i,\mathbf{x})),
\label{multi_eq2}
\end{eqnarray}
where, $\mathbf{x}=(x_1,x_2,\dots,x_N)^{\top}$, $G(x_i,\mathbf{x})=(G(x_i,x_1),\dots,G(x_i,x_N))^{\top}$, and  $\mathcal{L}(G(x_i,\mathbf{x}))$ represents the average input from the neighbors per unit edge weight under the heterogeneous meanfield approximation.
Under the approximation $\mathcal{L}(G(x_i,\mathbf{x}))\approx G(x_i,\mathcal{L}(\mathbf{x}))$, we can write Eq.~\eqref{multi_eq2} in vector form as 
\begin{eqnarray}
\frac{d\mathbf{x}}{dt}=F(\mathbf{x}) + \mathbf{s}^{\mathrm{in}}\circ G(\mathbf{x},\mathcal{L}(\mathbf{x})),\label{multi_eqvec}
\end{eqnarray}
where $G(\mathbf{x}, \mathcal{L}(\mathbf{x}))=(G(x_1,\mathcal{L}(\mathbf{x})),\dots,G(x_N,\mathcal{L}(\mathbf{x})))^{\top}$, $\mathbf{s}^{\mathrm{in}}=(s_1^{\mathrm{in}},\dots,s_N^{\mathrm{in}})^{\top}$,  and $\circ$ is the Hadamard product.
Because $\mathcal{L}$ is a linear operator, we obtain
\begin{eqnarray}
\frac{d\mathcal{L}(\mathbf{x})}{dt}&=&\mathcal{L}(F(\mathbf{x}) + \mathbf{s}^{\mathrm{in}}\circ G(\mathbf{x},\mathcal{L}(\mathbf{x})))\nonumber\\
&=&\mathcal{L}(F(\mathbf{x})) + \mathcal{L}(\mathbf{s}^{\mathrm{in}}\circ G(\mathbf{x},\mathcal{L}(\mathbf{x})))\nonumber\\
&\approx& F(\mathcal{L}(\mathbf{x})) + \mathcal{L}(\mathbf{s}^{\mathrm{in}}) G(\mathcal{L}(\mathbf{x}),\mathcal{L}(\mathbf{x})).
 \label{multi_eqvec2}
\end{eqnarray}
To derive the last line in Eq.~\eqref{multi_eqvec2}, we have assumed
$\mathcal{L}(F(\mathbf{x}))\approx F(\mathcal{L}(\mathbf{x}))$ and $\mathcal{L}(\mathbf{s}^{\mathrm{in}}\circ  G(\mathbf{x},\mathcal{L}(\mathbf{x})))\approx \mathcal{L}(\mathbf{s}^{\mathrm{in}})G(\mathcal{L}(\mathbf{x}),\mathcal{L}(\mathbf{x}))$.
 Gao {\it et al.} defined the effective state of the dynamical system by 
 \begin{eqnarray}
 x_{\mathrm{eff}}=\frac{\langle {s}^{\mathrm{out}}{x}\rangle}{\langle {s}^{\mathrm{out}}\rangle},
 \label{x_eff}
 \end{eqnarray}
which is the normalized weighted average of $x_i$ over all the nodes with  weight $s_i^{\mathrm{out}}$, and a control parameter
  \begin{eqnarray}
 \beta_{\mathrm{eff}}=\frac{\langle {s}^{\mathrm{in}}{s}^{\mathrm{out}}\rangle}{\langle {s}^{\mathrm{out}}\rangle}.
 \label{beta_eff}
  \end{eqnarray}
By substituting $\mathcal{L}(\mathbf{x})=x_{\mathrm{eff}}$ and $\mathcal{L}(\mathbf{s}^{\rm in})=\beta_{\mathrm{eff}}$ into Eq.~\eqref{multi_eqvec2}, we obtain  a one-dimensional reduction of the original $N$-dimensional dynamical system:
\begin{eqnarray}
\frac{dx}{dt}=F(x) +\beta G(x, x),
\label{1Dsystem}
\end{eqnarray}
which  is approximately satisfied by $(x,\beta)=(x_{\mathrm{eff}},\beta_{\mathrm{eff}})$. If the approximation is accurate, the equilibria and their stability of the one-dimensional dynamical system given by Eq.~\eqref{1Dsystem} can predict the resilience of the $N$-dimensional dynamical system given by Eq.~\eqref{multi_eq1}.
We can find the stable equilibria of the one-dimensional dynamical system given by Eq.~\eqref{1Dsystem}, which we denote by $x^*(\beta)$,   by setting its right-hand side to zero and excluding the unstable equilibria. The obtained  $x^*(\beta)$ is expected to approximate the effective state of the original $N$-dimensional dynamical system in a stable equilibrium, which we denote by $x^*_{\rm eff}$.

For Eq.~\eqref{1Dsystem} to accurately approximate the dynamics of $x_{\rm eff}$, the following approximations must hold accurately:
\begin{align} 
\text{Approximation (I) \ \ }: & \quad \mathcal{L}(F(\mathbf{x}))\approx F(\mathcal{L}(\mathbf{x})), 
\label{aprox1}\\
\text{Approximation (II) \ }: & \quad \mathcal{L}(G(x_i,\mathbf{x}))\approx G(x_i,\mathcal{L}(\mathbf{x})), 
\label{aprox2}\\
\text{Approximation (III)\;}: & \quad \mathcal{L}(\mathbf{s}^{\mathrm{in}}\circ  G(\mathbf{x},\mathcal{L}(\mathbf{x})))\nonumber\\ &\approx \mathcal{L}(\mathbf{s}^{\mathrm{in}})G(\mathcal{L}(\mathbf{x}),\mathcal{L}(\mathbf{x}))\,,
\label{aprox3}
\end{align}
and that (IV) the interaction network has negligible degree correlation.
For understanding when the GBB  reduction is accurate, we check the accuracy of approximations (I), (II), and (III) for different dynamical systems in Sec. \ref{section3},  assuming networks that satisfy condition (IV). The validity of approximations (I), (II), and (III) may depend on the dynamical system as well as the network structure. For example,  Eq.~\eqref{aprox3} does not hold true for diffusive coupling for the following reason. In the case of diffusive coupling, the right-hand side of Eq.~\eqref{aprox3} is always equal to zero because $G(\mathcal{L}(\mathbf{x}),\mathcal{L}(\mathbf{x}))=0$, whereas the left-hand side is not necessarily equal to zero because $\mathcal{L}(\mathbf{x})\neq x_i$ at least for some $x_i$ in general.

\section{Accuracy of the GBB reduction} \label{section3}

In this section we consider five types of dynamical systems and test the accuracy of the GBB  reduction and that of each of the three approximations used for deriving the one-dimensional reduction, i.e., (I), (II), and (III).  We use these dynamical systems because these dynamical systems were used in the previous studies related to the dimension reduction approach. Although this reduction approach is not applicable to the dynamical systems having a diffusive kind of coupling function. The reason behind this is  the right hand side of the equation \ref{aprox3} become zero for the diffusive coupling function, the left hand side is not necessarily zero. To understand the resilience of the  dynamical systems, we follow Ref.~\cite{Gao_Nature2016} to sequentially  remove nodes in a uniformly random order. After the removal of each node,  we also discard the nodes that do not belong to  the largest connected component and then compute $\beta_{\rm eff}$ using Eq.~\eqref{beta_eff} for the largest connected component.
We calculate the relative error given by
\begin{equation}
\mathrm{Relative~error} = \left| \frac{x^*(\beta_{\mathrm{eff}}) - x^*_{\mathrm{eff}}}{x^*(\beta_{\mathrm{eff}})} \right|\,.
\label{error}
\end{equation}
We obtain $x^*_{\rm eff}$ from the stable equilibrium of the original $N$-dimensional system. We calculate $x^*_{\rm eff}$ by running Eq.~\eqref{multi_eq1}  using the fourth-order Runge-Kutta method (with MATLAB function ode45) until it converges and substituting the final values of $x_i$ (with $i=1,\dots,N$), which we denote by $x_i^*$, into Eq.~\eqref{x_eff}. When there are multiple stable equilibria of Eq.~\eqref{1Dsystem},
we calculate the relative error with respect to the stable equilibrium $x^*(\beta_{\rm eff})$ that is the closest to $x^*_{\rm eff}$.

To investigate the accuracy of approximations (I), (II), and (III), we calculate the ratio of the right-hand side to the left-hand side of Eqs.~\eqref{aprox1}, \eqref{aprox2}, and \eqref{aprox3}, which we refer to as
\begin{align}
R^{(1)} =& \frac{ \mathcal{L}\left(\mathbf{s}^{\mathrm{in}}\right)G\left(\mathcal{L}(\mathbf{x}),\mathcal{L}\left(\mathbf{x}\right)\right)}{\mathcal{L}\left(\mathbf{s}^{\mathrm{in}}\circ  G\left(\mathbf{x},\mathcal{L}(\mathbf{x})\right)\right)}\,,\\
R^{(2)} =& \frac{G\left(x_i,\mathcal{L}\left(\mathbf{x}\right)\right)}{\mathcal{L}\left(G\left(x_i,\mathbf{x}\right)\right)}\,,
\end{align}
and
\begin{align}
R^{(3)} =& \frac{F\left(\mathcal{L}\left(\mathbf{x}\right)\right)}{\mathcal{L}\left(F\left(\mathbf{x}\right)\right)}\,,
\end{align}
respectively. The GBB  reduction is exact for uncorrelated networks when all of $R^{(1)}$, $R^{(2)}$, and $R^{(3)}$ are equal to 1.  Therefore, even for the uncorrelated networks, the accuracy of the GBB  reduction is not guaranteed and the accuracy depends on the considered dynamical systems' intrinsic dynamics and the coupling function.

For numerical verification, we use two undirected and unweighted networks with $N=500$ nodes, i.e., a regular random graph with node's degree 8 and a scale-free network of approximately the same average degree (i.e., $\langle k \rangle=7.96$). We construct the scale-free network using the Barab{\'a}si-Albert model \cite{Barabasi_Science1999} with the number of edges per new node $m=4$ and the initial condition of the complete graph of four nodes. The model produces the degree distribution $P(k) \propto k^{-3}$ for large $k$. We use these networks because they do not have degree correlation and therefore a high accuracy of approximation (IV) is expected.  If we use correlated networks, such as empirical networks, model networks with community structure or clustering, then we would not be able to dissociate the reason for the approximation error. This is because the error, if nonnegligible, may be due to the correlation in the network structure or the due to the assumptions underlying the GBB  reduction that are not guaranteed to be accurate even for uncorrelated networks.

\subsection{Double-well system} \label{double_well}

We first consider a coupled double-well potential system given by
\begin{equation}
\frac{dx_i}{dt}=-\left(x_i-r_1\right)\left(x_i-r_2\right)\left(x_i-r_3\right) +D\sum_{j=1}^N A_{ij}x_j, 
\label{double_well_system1}
\end{equation}
where $x_i$ is the state of the $i$th node, and $D$ is the coupling strength \cite{Wunderling_Chaos2020, Brummitt_JRSI2015, Kronke_PRE2020}. The model for a single node is called the Schl{\"o}gl model \cite{Schlogl_ZeiPhy1972, Kouvaris_PlosOne2012}.  We assume  $r_1<r_2<r_3$ without loss of generality. In the absence of the coupling term, this dynamical system  has two stable equilibria, $x=r_1$ and $x=r_3$. 
The GBB  reduction is given by
\begin{equation}
\frac{dx}{dt}=-\left(x-r_1\right)\left(x-r_2\right)\left(x-r_3\right) +D \beta x\,.
\label{double_well_system1D}
\end{equation}
\begin{figure}
\centering
\includegraphics[height=!,width=0.9\textwidth]{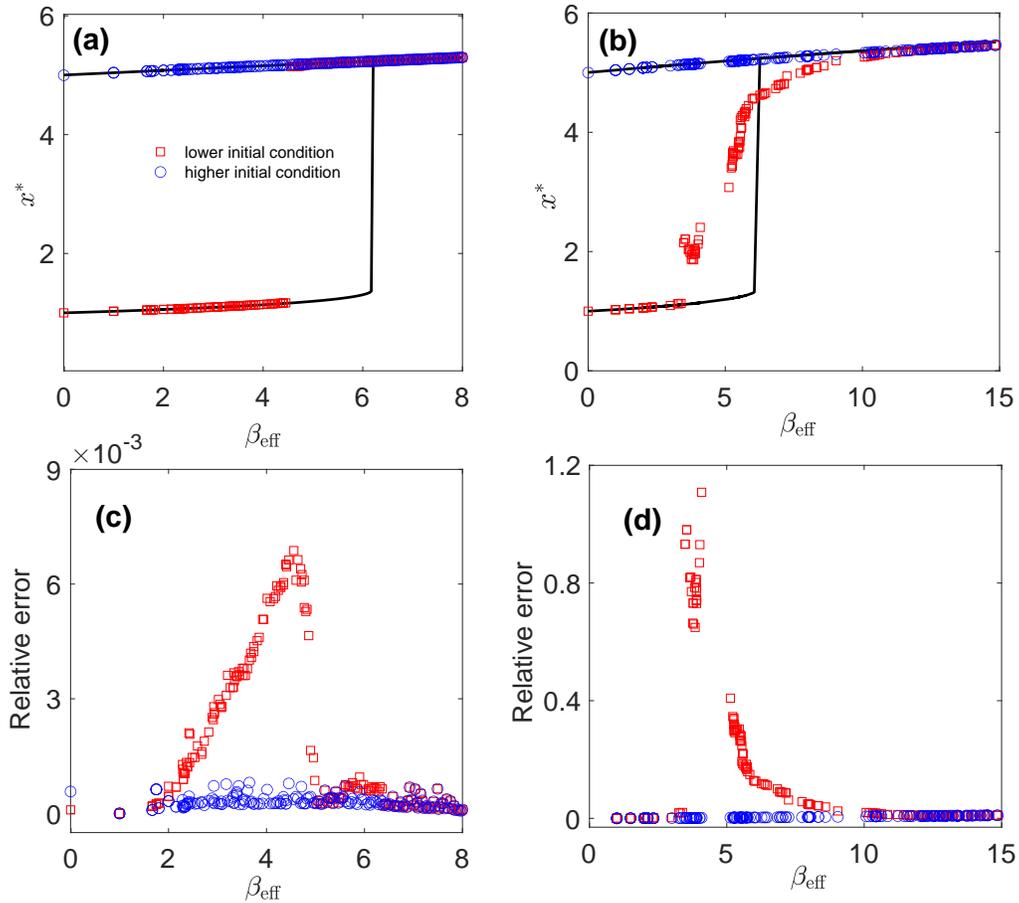}
\caption{{\bf GBB  reduction for the double-well system.} (a) Bifurcation diagram for the regular random graph. (b) Bifurcation diagram for the scale-free network. (c) Relative error for the regular random graph. (d) Relative error for the scale-free network. The squares and circles represent the numerically obtained equilibria when the initial condition is $x_i=0.01$ for all $i$ and $x_i=10$ for all $i$, respectively. At each $\beta_{\mathrm{eff}}$ value, we started the simulation of the original dynamical system from each of these two initial conditions and obtained the equilibria. The solid lines in (a) and (b) represent the stable equilibria of the GBB  reduction given by Eq.~\eqref{double_well_system1D}.} 
\label{double_well_fig1}
\end{figure}
\begin{figure}
\centering
\includegraphics[height=!,width=0.9\textwidth]{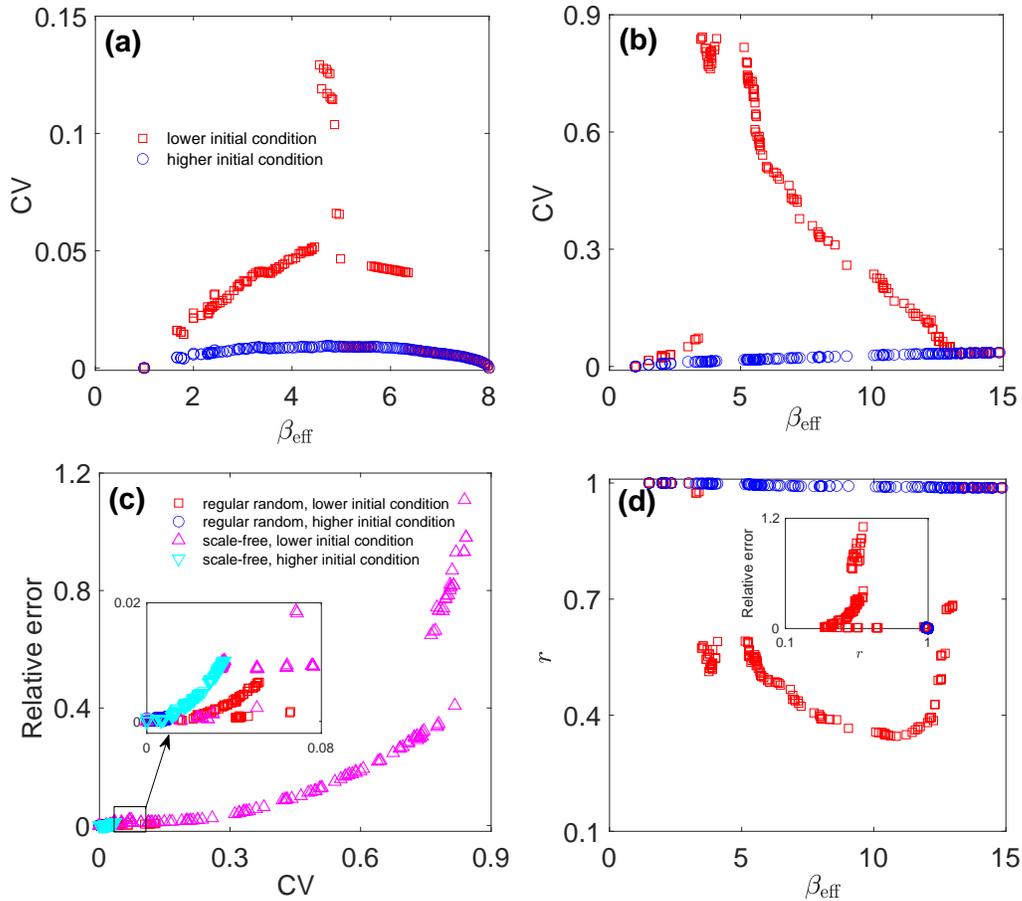}
\caption{{\bf Exploring reasons for the accuracy of the GBB  reduction for the double-well system.} (a) CV of $\{x^*_1, \ldots, x^*_N\}$ as a function of $\beta_{\rm eff}$ for the regular random graph. (b) CV of $\{x^*_1, \ldots, x^*_N\}$ as a function of $\beta_{\rm eff}$  for the scale-free network.  (c) Relative error as a function of the CV for both  networks. The squares and circles correspond to the lower and higher initial conditions, respectively, for the regular random graph. The triangles and inverted triangles correspond to the lower and higher initial conditions, respectively, for the scale-free network. (d) Pearson correlation coefficient, denoted by $r$, between $x^*_i$ and $k_i$  for the scale-free network. The inset shows the relationship between the relative error and $r$. In (a), (b), and (d), the squares and circles represent the lower and higher initial conditions, respectively.} 
\label{double_well_fig2}
\end{figure}
We show the relationship between $x^*_{\rm eff}$ and $\beta_{\rm eff}$ for the regular random graph and scale-free network in Figs.~\ref{double_well_fig1}(a) and \ref{double_well_fig1}(b), respectively, both for the original dynamical system (squares and circles; Eq.~\eqref{double_well_system1}) and its one-dimensional reduction (solid lines; $x^*(\beta_{\text{eff}})$ obtained from Eq.~\eqref{double_well_system1D}).  
We set $r_1=1$, $r_2=2$, $r_3=5$, and $D=0.1$ in this and the following numerical simulations in this section. The relative error corresponding to Figs.~\ref{double_well_fig1}(a) and \ref{double_well_fig1}(b) is shown in Figs.~\ref{double_well_fig1}(c) and \ref{double_well_fig1}(d), respectively. The relative error is small except near the bifurcation point. The GBB  reduction is inaccurate at estimating the location of the bifurcation point in terms of $\beta_{\text{eff}}$. 

For this dynamical system, we obtain $F\left(x_i\right)=-\left(x_i-r_1\right)\left(x_i-r_2\right)\left(x_i-r_3\right)$ and $G\left(x_i,x_j\right)=x_j$.
Because $G(x_i,x_j)$ is linear in terms of $x_j$, we obtain $R^{(2)} = R^{(3)} = 1$. In other words, Eqs.~\eqref{aprox2} and \eqref{aprox3} hold true with equality. Therefore, we only focus on examining the accuracy of the approximation given by Eq.~\eqref{aprox1}.

We obtain
\begin{widetext}
\begin{eqnarray}
R^{(1)}&=&\frac{F\left(\mathcal{L}\left(\mathbf{x}\right)\right)}{\mathcal{L}\left(F\left(\mathbf{x}\right)\right)}\nonumber\\
&=&\frac{ \left(\frac{\langle s^{\mathrm{out}}{x}\rangle}{\langle s^{\mathrm{out}}\rangle}-r_1\right)\left( \frac{\langle s^{\mathrm{out}}{x}\rangle}{\langle s^{\mathrm{out}}\rangle}-r_2\right)\left(\frac{\langle s^{\mathrm{out}}{x}\rangle}{\langle s^{\mathrm{out}}\rangle}-r_3\right)}{\frac{\langle s^{\mathrm{out}}\left({x}-r_1\right)\left({x}-r_2\right)\left({x}-r_3\right)\rangle }{\langle s^{\mathrm{out}}\rangle} }\nonumber\\
&=&\frac{ {\langle s^{\mathrm{out}}{x}\rangle}^3-  \left(r_1+r_2+r_3\right){\langle s^{\mathrm{out}}{x}\rangle}^2\langle s^{\mathrm{out}}\rangle +  \left(r_1r_2  +r_2r_3+r_3r_1\right){\langle s^{\mathrm{out}}{x}\rangle}\langle s^{\mathrm{out}}\rangle^2-r_1r_2r_3\langle s^{\mathrm{out}}\rangle^3} {\langle s^{\mathrm{out}}\rangle^2\langle s^{\mathrm{out}}{x}^3\rangle - \left(r_1+r_2+r_3\right)\langle{s}^{\mathrm{out}}{x}^2\rangle \langle s^{\mathrm{out}}\rangle^2 +\left(r_1r_2  +r_2r_3+r_3r_1\right)\langle{s}^{\mathrm{out}}{x} \rangle \langle s^{\mathrm{out}}\rangle^2- r_1r_2r_3  \langle s^{\mathrm{out}}\rangle^3} .\nonumber \\
\label{dweq12}
\end{eqnarray}
\end{widetext}

The last two terms in the numerator and denominator of Eq.~\eqref{dweq12} are the same. Therefore the inaccuracy of the GBB  reduction results from the discrepancy between the first two terms of the numerator and those of the denominator. 
The approximation is exact for uncorrelated networks (i.e., $R^{(1)} = 1$) when all $x_i$'s are the same such that $\langle {x}^2\rangle / \langle {x}\rangle^2 = \langle {x}^3\rangle / \langle {x}\rangle^3 =1$ or when $s^{\rm out}$ is independent of $x$.  

To numerically examine these quantities,  we plot the coefficient of variation (CV), which is the standard deviation divided by the average of $x^*_i$ (with $i=1, \ldots, N$), given by 
\begin{eqnarray}
{\rm CV} =\frac{\sqrt{\frac{1}{N}\left(\sum_{i=1}^N(x^*_i -\langle x^* \rangle\right)^2}}{\langle x^*\rangle}
\end{eqnarray}
for the regular random graph and scale-free network in Figs.~\ref{double_well_fig2}(a) and \ref{double_well_fig2}(b), respectively. We find that the CV is small when the relative error (see Figs.~ \ref{double_well_fig1}(c) and  \ref{double_well_fig1}(d)) is small, which is consistent with the fact that there is no approximation error when all the $x^*_i$ values are the same (which is equivalent to CV $= 0$). We plot the relative error as a function of the CV in Fig.~\ref{double_well_fig2}(c).
Figure \ref{double_well_fig2}(c) shows that the relation between the CV and the relative error is largely similar between the two networks and that the magnitude of the error increases as the CV increases. These results are consistent with the observation that, given the $\beta_{\rm eff}$ value, the CV (see Figs.~\ref{double_well_fig2}(a) and \ref{double_well_fig2}(b)) and the relative error (see Figs.~\ref{double_well_fig1}(c) and \ref{double_well_fig1}(d)) are generally smaller for the regular random graph than the scale-free network.

Because the independence of $x^*_i$ and $k_i (\equiv s_i^{\rm in} = s_i^{\rm out})$, where $k_i$ is the degree of the $i$th node, is another sufficient condition for $R^{(1)} = 1$, we also examine the Pearson correlation coefficient  between $x^*_i$ and $k_i$, denoted by $r$, for the scale-free network. For the regular random graph, we do not calculate $r$ because all the nodes have the same $k_i$ value in the original network and the dispersion of $k_i$ remains small after removals of uniformly randomly selected nodes. The $r$ values shown in Fig.~\ref{double_well_fig2}(d) indicate that the correlation is  almost equal to 1 for any $\beta_{\mathrm{eff}}$ in the case of the higher  equilibrium. For the lower  equilibrium, the dependence of $r$ on $\beta_{\mathrm{eff}}$  is not consistent with the dependence of the relative error on $\beta_{\mathrm{eff}}$ (the squares in Fig.~\ref{double_well_fig1}(d); also see the inset). Furthermore, the relative error is negligible for the lower equilibrium anyway. 

Therefore, we conclude that the CV of $\{x^*_1,\dots,x^*_N\}$, not the correlation between  $x^*_i$ and $k_i$, is  a major determinant of the accuracy of the GBB  reduction for the double-well system. 

\subsection{SIS model} \label{SIS_section}
Next, we consider the deterministic approximation to the stochastic susceptible-infectious-susceptible (SIS) dynamics, which is also called the individual-based approximation \cite{Pastor_RMP2015}. It is given by 
\begin{eqnarray}
\frac{dx_i}{dt}= \lambda\sum_{j=1}^N A_{ij}(1-x_i)x_j-\mu x_i , 
\label{SISeq1}
\end{eqnarray}
where $x_i$ represents the probability that the $i$th node is  infectious at time $t$,  $\lambda $ is the infection rate, and $\mu$ is the recovery rate.  The first term on the right-hand side of Eq.~\eqref{SISeq1} represents the rate at which the $i$th node is infected by one of its neighbors. The second term represents the recovery. The GBB  reduction for Eq.~\eqref{SISeq1} is given by
\begin{eqnarray}
\frac{dx}{dt}= \lambda\beta\left(1-x\right)x-\mu x\,.
\label{SISeq1D}
\end{eqnarray}

\begin{figure}
\includegraphics[height=!,width=0.9\textwidth]{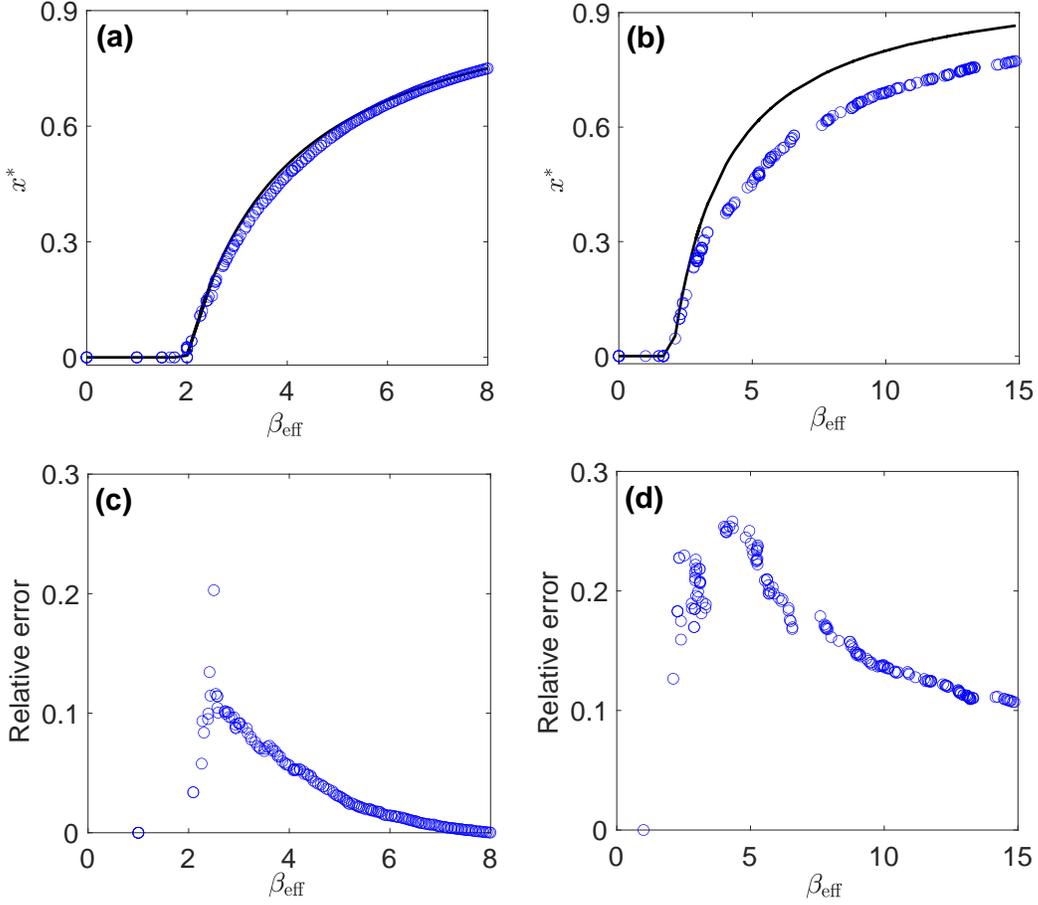}
\caption{{\bf GBB  reduction for the SIS model. }(a) Bifurcation diagram  for the regular random graph. (b) Bifurcation diagram  for the scale-free network. (c) Relative error for the regular random graph.  (d) Relative error  for the scale-free network. The circles represent the numerically obtained equilibria with the initial condition $x_i= 10$ for all $i$. The solid curves in (a) and (b) represent the stable equilibria of the GBB  reduction.
We set $\lambda = 0.5$ and $\mu = 1$.} 
\label{SISFigure1}
\end{figure}

In Figs.~\ref{SISFigure1}(a) and \ref{SISFigure1}(b), we show the relation between $x^*_{\rm eff}$ and $\beta_{\rm eff}$ for the regular random graph and scale-free network, respectively, each for the original dynamical system given by Eq.~\eqref{SISeq1} and the GBB  reduction given by Eq.~\eqref{SISeq1D} (i.e., $x^*(\beta_{\mathrm{eff}})$). We set $\lambda=0.5$ and  $\mu=1$. We find that the GBB  reduction accurately estimates the location of the epidemic threshold in terms of $\beta_{\mathrm{eff}}$ for both networks. We show the relative error of the GBB  reduction for the regular random graph and scale-free network in Figs.~\ref{SISFigure1}(c) and \ref{SISFigure1}(d), respectively. These figures indicate
 that the error is larger for the scale-free network  than the regular random graph.
 
For the SIS model, we have $F(x_i)=-\mu x_i$ and $G(x_i,x_j)=\lambda (1-x_i)x_j$. Therefore, Eq.~\eqref{aprox1} holds with equality (i.e., $R^{(1)} = 1$) for any $\mathbf{x}$ because $F(x_i)$ is linear in $x_i$. Equation \eqref{aprox2} also holds with equality (i.e., $R^{(2)} = 1$) because
\begin{eqnarray}
\mathcal{L}\left(G\left(x_i,\mathbf{x}\right)\right)&=&\lambda \mathcal{L}\left(\left(1-x_i\right)\mathbf{x} \right) \nonumber\\
&=&\lambda\left(1-x_i\right)\frac{\langle {s}^{\mathrm{out}}{x}\rangle}{\langle {s}^{\mathrm{out}}\rangle}  \nonumber\\
&=&G\left(x_i,\mathcal{L}\left(\mathbf{x}\right)\right).
\end{eqnarray}
For the third approximation (see Eq.~\eqref{aprox3}), we obtain 
\begin{eqnarray}
R^{(3)}&=&\frac{\mathcal{L}\left(\mathbf{s}^{\mathrm{in}}\right)G\left(\mathcal{L}\left(\mathbf{x}\right),\mathcal{L}\left(\mathbf{x}\right)\right)}{\mathcal{L}\left(\mathbf{s}^{\mathrm{in}}\circ  G\left(\mathbf{x},\mathcal{L}\left(\mathbf{x}\right)\right)\right)} \nonumber\\
&=&\frac{ \frac{\langle {s}^{\mathrm{in}}{s}^{\mathrm{out}}\rangle}{\langle {s}^{\mathrm{out}}\rangle} \lambda \left(1-\frac{\langle {s}^{\mathrm{out}}{x}\rangle}{\langle {s}^{\mathrm{out}}\rangle}\right) \frac{\langle {s}^{\mathrm{out}}{x}\rangle}{\langle {s}^{\mathrm{out}}\rangle}} {\lambda\frac{\left(\langle {s}^{\mathrm{in}}{s}^{\mathrm{out}}\rangle-\langle {s}^{\mathrm{in}}{s}^{\mathrm{out}} {x}\rangle\right)\langle {s}^{\mathrm{out}}{x}\rangle}{\langle {s}^{\mathrm{out}}\rangle^2}  } \nonumber\\
&=&\frac{ {\langle {s}^{\mathrm{in}}{s}^{\mathrm{out}}\rangle} \left(1-\frac{\langle {s}^{\mathrm{out}}{x}\rangle}{\langle {s}^{\mathrm{out}}\rangle}\right) } { \langle {s}^{\mathrm{in}}{s}^{\mathrm{out}}\rangle-\langle {s}^{\mathrm{in}}{s}^{\mathrm{out}} {x}\rangle } \nonumber\\
&=&\frac{  1-\frac{\langle {s}^{\mathrm{out}}{x}\rangle}{\langle {s}^{\mathrm{out}}\rangle} } {1-\frac{\langle {s}^{\mathrm{in}}{s}^{\mathrm{out}} {x}\rangle}{{\langle {s}^{\mathrm{in}}{s}^{\mathrm{out}}\rangle}} } .
\label{SISap1}
\end{eqnarray}
Therefore, the GBB  reduction is accurate if
\begin{eqnarray}
\frac{\langle {s}^{\mathrm{out}}{x}\rangle}{\langle {s}^{\mathrm{out}}\rangle} \approx \frac{\langle {s}^{\mathrm{in}}{s}^{\mathrm{out}} {x}\rangle}{\langle {s}^{\mathrm{in}}{s}^{\mathrm{out}}\rangle}.
\label{terms}
\end{eqnarray}

\begin{figure}
\includegraphics[height=!,width=0.9\textwidth]{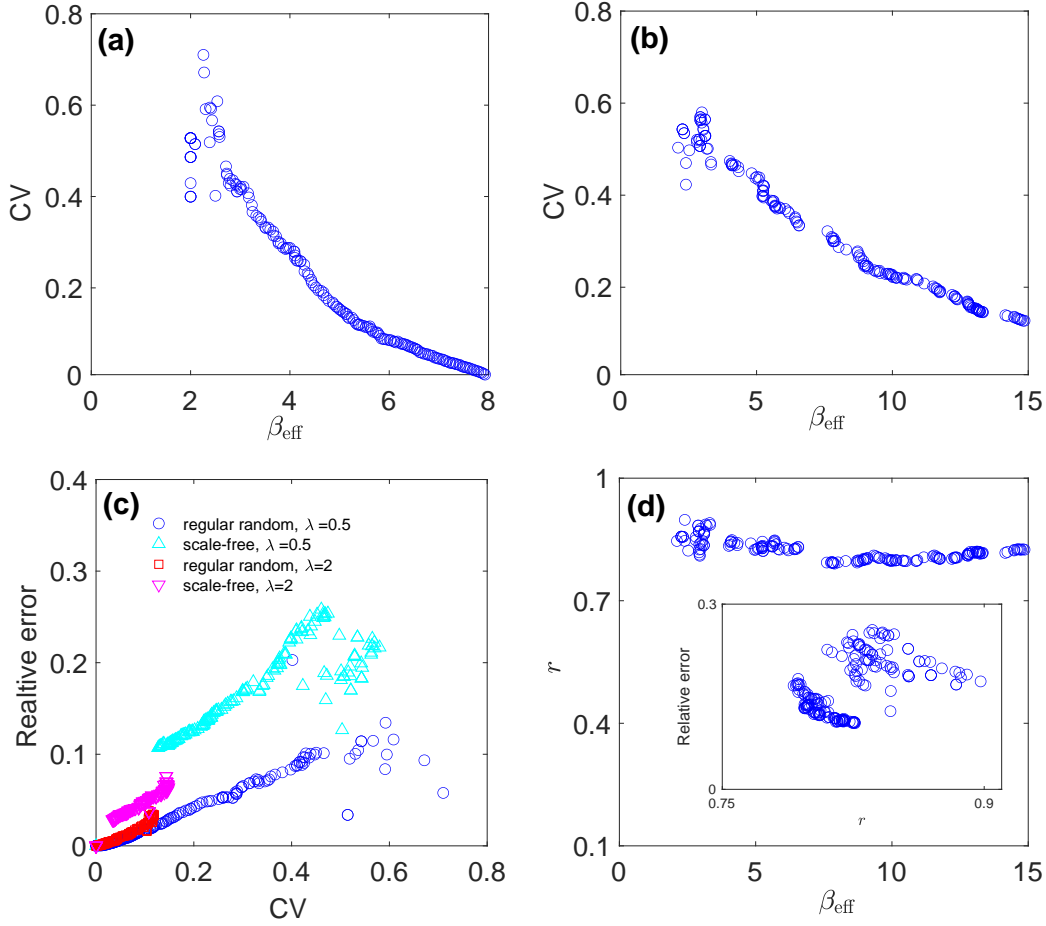}
\caption{{\bf Exploring reasons for the accuracy of the GBB  reduction for the SIS model.} (a) CV of $\{x^*_1, \ldots, x^*_N\}$ for the regular random graph. (b) CV of $\{x^*_1, \ldots, x^*_N\}$ for the scale-free network. (c)  Relative error as a function of the CV for the two networks and two values of $\lambda$, i.e., $\lambda = 0.5$ and $\lambda=2$. (d) Pearson correlation coefficient between the state variable and the out-degree, $r$, as a function of $\beta_{\rm eff}$ for the scale-free network. The inset shows the relationship between the relative error and $r$.
We set $\mu = 1$. In (a), (b), and (d), we set $\lambda = 0.5$.} 
\label{SISFigure2}
\end{figure}

Equation \eqref{terms} holds with equality when $x_i$ is independent of $i$ or $x_i$ is independent of $s_i^{\rm out}$ and $s_i^{\rm in}s_i^{\rm out}$. In the case of the regular random graph, $s_i^{\rm in}$ and $s_i^{\rm out}$ are the same for all the nodes such that the second condition is satisfied. To examine the first possibility, we plot the CV of $\{x^*_1, \ldots, x^*_N\}$ as a function of $\beta_{\mathrm{eff}}$ in Figs.~\ref{SISFigure2}(a) and  \ref{SISFigure2}(b) for the regular random graph and scale-free network, respectively. 
For both networks, the CV largely decreases as $\beta_{\rm eff}$ increases. 
In Fig.~\ref{SISFigure2}(c), we show the relationships between the relative error and the CV for the two networks and two values of $\lambda$ (i.e., $\lambda=0.5$ and $\lambda=2$). The figure indicates that, for both values of $\lambda$,
the relative error increases as the CV increases except at large CV values. These results are consistent with the fact that the approximation error is zero when the $x_i$ value is independent of $i$.
Furthermore, Fig.~\ref{SISFigure2}(c) indicates that the relation between the relative error and the CV is quantitatively close, albeit not the same, between the two networks given the $\lambda$ value.
%

We show in Fig.~\ref{SISFigure2}(d) the Pearson correlation coefficient, $r$, between $x^*_i$ and $k_i$  as a function of $\beta_{\mathrm{eff}}$ for the scale-free network. The correlation coefficient $r$ is large (i.e., $>0.75$) across the range of  $\beta_{\mathrm{eff}}$ including the values of $\beta_{\mathrm{eff}}$ for which the relative error is small (see Fig.~\ref{SISFigure1}(d)). Furthermore, the relative error and $r$ are apparently unrelated (see the inset of Fig.~\ref{SISFigure2}(d)). Therefore, we conclude that $r$ is not a determinant of the accuracy of GBB  reduction and that the CV of $\{x^*_1,\dots,x^*_N\}$ heavily impacts the accuracy of the GBB  reduction. These conclusions are similar to those for the double-well system.

\subsection{Gene regulatory system} \label{gene}

In this section, we consider a model of a gene regulatory system governed by the Michaelis-Menten equation  \cite{Gao_Nature2016}, which is given by 
\begin{eqnarray}
\frac{dx_i}{dt}=-Bx_i^f +\sum_{j=1}^N A_{ij}\frac{x_j^h}{1+x_j^h}\,,
\label{R_system1}
\end{eqnarray}
where $x_i$ represents the expression level of gene $i$. The first term on the right-hand side of  Eq.~\eqref{R_system1} represents the degradation. The second term represents the activation of gene $i$ by gene $j$. We use the same parameter values as those in Ref.~\cite{Gao_Nature2016}, i.e., $B=1$, $f=1$, and $h=2$.
The GBB  reduction corresponding to Eq.~\eqref{R_system1} is given by
\begin{eqnarray}
\frac{dx}{dt}=-Bx^f +\beta \frac{x^h}{1+x^h}.
\label{R_system1D}
\end{eqnarray}

\begin{figure}
\includegraphics[height=!,width=0.9\textwidth]{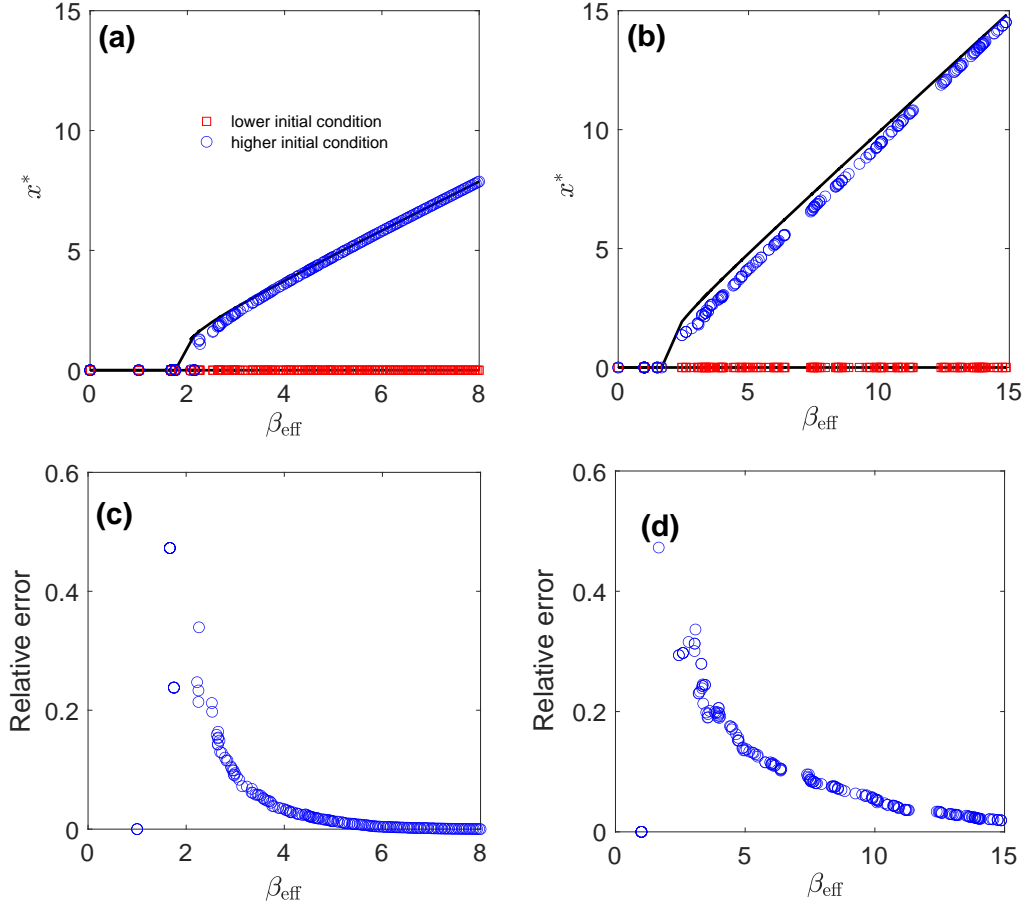}
\caption{{\bf GBB  reduction for the gene regulatory system.} (a) Bifurcation diagram  for the regular random graph. (b) Bifurcation diagram for the scale-free network. (c) Relative error at the nontrivial equilibria for the regular random graph. (d) Relative error at the nontrivial equilibria for the scale-free network. The circles and squares represent the numerically obtained equilibria  with the initial conditions $x_i= 10$ and $x_i= 0.01$, respectively, for all $i$.} 
\label{Figure21}
\end{figure}

We show the bifurcation diagrams for the original dynamical system (i.e., Eq.~\eqref{R_system1}) and its  GBB  reduction (i.e., Eq.~\eqref{R_system1D}) in Figs.~\ref{Figure21}(a) and \ref{Figure21}(b) for the regular random graph and scale-free network, respectively. 
We plot the relative error  of the GBB  reduction at the nontrivial equilibria in Figs.~\ref{Figure21}(c) and  \ref{Figure21}(d) for the regular random graph and scale-free network, respectively.  For both networks, the relative error is small except near the bifurcation point.

Equation \eqref{R_system1}  yields 
\begin{eqnarray}
F\left(x_i\right)=Bx_i
\label{R_systemFx}
\end{eqnarray}
and 
\begin{eqnarray}
G\left(x_i,x_j\right)=\frac{x_j^2}{1+x_j^2}\,.
\label{R_systemgx}
\end{eqnarray}
The approximation in Eq.~\eqref{aprox1} is exact for this dynamical system (i.e., $R^{(1)} = 1$) because $F\left(x_i\right)$ is linear in $x_i$. We then obtain 

\begin{figure}
\centering
\includegraphics[height=!,width=0.9\textwidth]{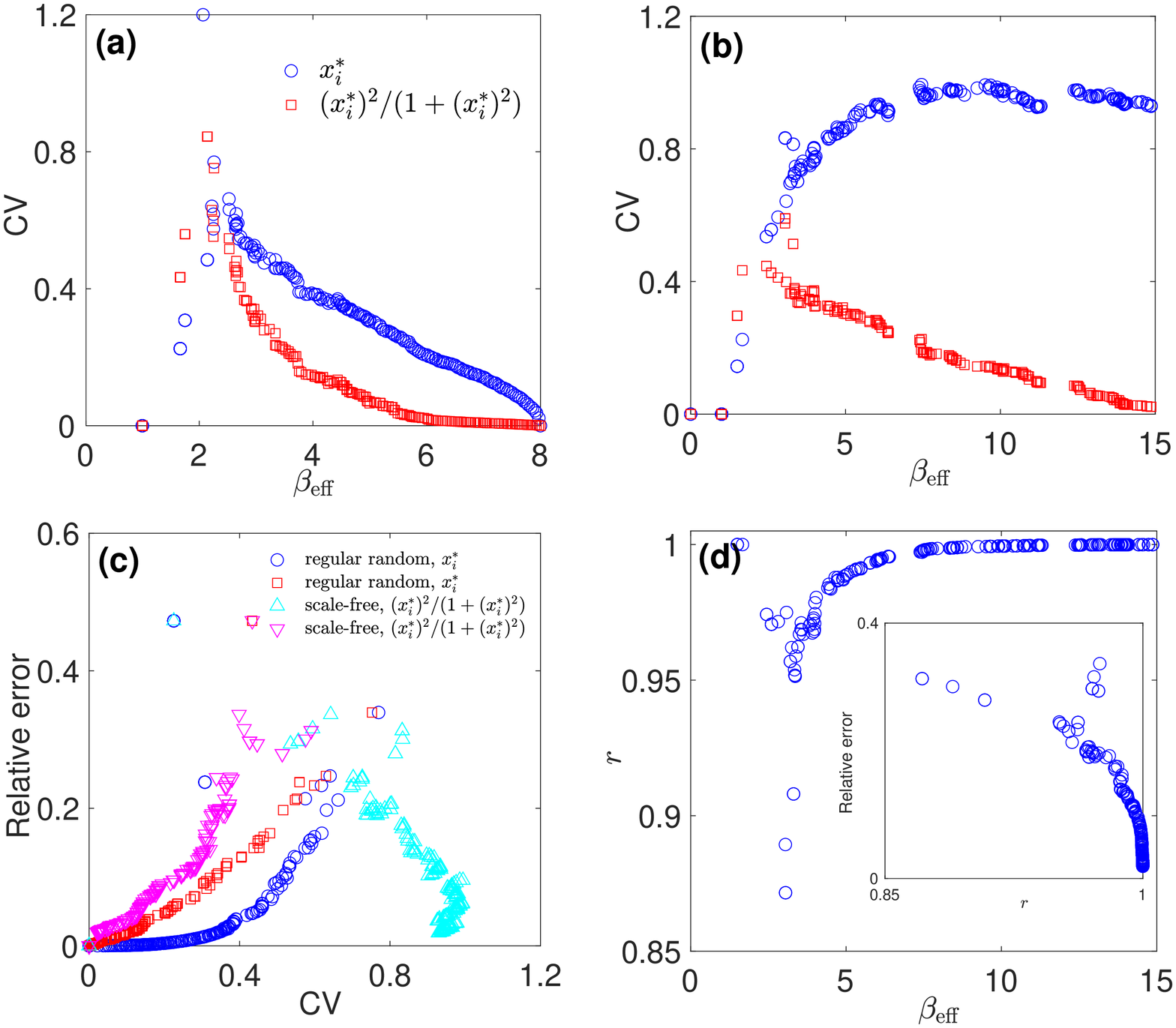}
\caption{{\bf Accuracy of the GBB  reduction for the gene regulatory system.} (a) CV of $\{x^*_1,\ldots,x^*_N\}$ and of $\left\{(x^*_1)^2/(1+(x^*_1)^2), \ldots, (x^*_N)^2/(1+(x^*_N)^2) \right\}$ as a function of $\beta_{\mathrm{eff}}$ for the regular random graph. 
(b) CV of $\{x^*_1, \dots, x^*_N\}$ and of $\left\{ (x^*_1)^2/(1+(x^*_1)^2), \ldots, (x^*_N)^2/(1+(x^*_N)^2) \right\}$ as a function of  $\beta_{\mathrm{eff}}$ for the scale-free network. The circles and squares correspond to the CV of $\{x^*_1, \ldots, x^*_N\}$ and that of $\left\{ (x^*_1)^2/(1+(x^*_1)^2), \ldots, (x^*_N)^2/(1+(x^*_N)^2) \right\}$, respectively. (c) Relative error as a function of the CV for the two networks. The circles and triangles correspond to the CV of $\{x^*_1, \ldots, x^*_N\}$ for the regular random graph and scale-free network, respectively. The squares and inverted triangles correspond to the CV of $\left\{ (x^*_1)^2/(1+(x^*_1)^2), \ldots, (x^*_N)^2/(1+(x^*_N)^2) \right\}$ for the regular random graph and scale-free network, respectively.
(d) Pearson correlation coefficient between $x^*_i$ and $k_i$, i.e., $r$, as a function of  $\beta_{\mathrm{eff}}$ for the scale-free network. 
The inset shows the relationship between the relative error and $r$.} 
\label{Figure22}
\end{figure}

\begin{eqnarray}
R^{(2)}&=&\frac{G\left(x_i,\mathcal{L}\left(\mathbf{x}\right)\right)}{\mathcal{L}\left(G\left(x_i,\mathbf{x}\right)\right)}\nonumber\\
&=& \frac{\langle {s}^{\mathrm{out}}{x}\rangle ^2\langle {s}^{\mathrm{out}}\rangle}{\left(\langle {s}^{\mathrm{out}}\rangle ^2 +\langle {s}^{\mathrm{out}}{x}\rangle ^2\right)\langle {s}^{\mathrm{out}}\frac{{x^2}}{1+{x^2}}\rangle} \,.
\label{gene_aprox2}
\end{eqnarray}
Finally, the approximation given by Eq.~\eqref{aprox3} is exact (i.e., $R^{(3)} = 1$) because
\begin{eqnarray}
\mathcal{L}\left(\mathbf{s}^{\mathrm{in}}\circ  G\left(\mathbf{x},\mathcal{L}\left(\mathbf{x}\right)\right)\right)&=&\mathcal{L} \left(\mathbf{s}^{\mathrm{in}}\circ\frac{\left(\mathcal{L}\left(\mathbf{x}\right)\right)^2 }{1+\left(\mathcal{L}\left(\mathbf{x}\right)\right)^2}\right) \nonumber \\
&=&\frac{\left(\mathcal{L}\left(\mathbf{x}\right)\right)^2 }{1+\left(\mathcal{L}\left(\mathbf{x}\right)\right)^2}\mathcal{L} \left(\mathbf{s}^{\mathrm{in}}\circ\mathbf{1}\right) \nonumber \\
&=&\mathcal{L}\left(\mathbf{s}^{\mathrm{in}}\right)G\left(\mathcal{L}\left(\mathbf{x}\right),\mathcal{L}\left(\mathbf{x}\right)\right)\,.
\end{eqnarray}
Therefore, the inaccuracy of the GBB  reduction in the case of uncorrelated networks only originates from the deviation of  $R^{(2)}$ from 1.
Equation~\eqref{gene_aprox2} implies that the approximation holds without error when $x_i$ is independent of $i$ or $s_i^{\rm out}$ is independent of $x_i$ and $x_i^2/(1+x_i^2)$. 

We plot the CV of $\{x^*_1,\dots,x^*_N\}$  as a function of $\beta_{\rm eff}$ by the circles in Figs.~\ref{Figure22}(a) and \ref{Figure22}(b) for the regular random graph and scale-free network, respectively. The CV of $\{x^*_1,\dots,x^*_N\}$  decreases with the increase in $\beta_{\rm eff}$ for both networks. We plot the relative error against the CV for the regular random graph and scale-free network by the circles and triangles in Fig.~\ref{Figure22}(c), respectively. The figure shows that, although the relative error largely increases as the CV increases for both networks, the two networks show considerably different relationships between the relative error and the CV. Therefore, we consider that the CV of $\{x^*_1,\dots, x^*_N\}$ does not explain the magnitude of the relative error sufficiently well. Equation~\eqref{R_system1} with $h=2$ suggests that the interaction term saturates according to $x_j^2/(1+x_j^2)$ as $x_j$ increases. The squares in Figs.~\ref{Figure22}(a) and \ref{Figure22}(b) represent the CV of $\left\{ (x^*_1)^2/(1+(x^*_1)^2), \ldots, (x^*_N)^2/(1+(x^*_N)^2) \right\}$ as a function of $\beta_{\rm eff}$ for the regular random graph and scale-free network, respectively. We find that the CV decreases as $\beta_{\rm eff}$ increases in both networks except at small $\beta_{\rm eff}$ values. We show the relative error as a function of the CV of $\left\{(x^*_1)^2/(1+(x^*_1)^2), \ldots, (x^*_N)^2/(1+(x^*_N)^2) \right\}$ for the regular random graph and scale-free network by the squares and inverted triangles, respectively, in Fig.~\ref{Figure22}(c). We find that the dependence of the relative error on the CV is now much more similar between the two networks than in the case of the dependence of the relative error on the CV of $\{ x^*_1, \ldots, x^*_N \}$.
The Pearson correlation coefficient between $x_i$ and $k_i$ for the scale-free network, denoted by $r$, is shown in Fig.~ \ref{Figure22}(d). The figure shows that $r$ is large regardless of the $\beta_{\rm eff}$ value and that the relative error decreases as $r$ increases.
The latter result is difficult to reconcile because, as we stated above, the independence between $s_i^{\rm out}$ and $x_i$ (i.e., $r=0$) is part of a sufficient condition for the GBB  reduction to be exact, which would yield the zero relative error; the inset of Fig.~ \ref{Figure22}(d) shows an opposite tendency.
Therefore, we conclude that $r$ does not explain the accuracy of the GBB  reduction. 
In sum, in the case of this gene regulatory system, the CV of $\left\{ (x^*_1)^2/(1+(x^*_1)^2), \ldots, (x^*_N)^2/(1+(x^*_N)^2) \right\}$, rather than that of $\{x^*_1, \ldots, x^*_N\}$ or the correlation between $x^*_i$ and $k_i$, explains  the accuracy of the GBB  reduction.

\subsection{Generalized Lotka-Volterra model} \label{GLV_section}

The generalized Lotka-Volterra (GLV) dynamics is given by
\begin{eqnarray}
\frac{dx_i}{dt}=\alpha x_i +\sum_{j=1}^N A_{ij}x_ix_j,
\label{GLVeq1}
\end{eqnarray}
where $x_i$ represents the abundance of the $i$th species, $\alpha$ is the intrinsic growth rate of the species, and $N$ is the  number of species in the population \cite{Tu_PRE2017}. The nontrivial equilibrium of this dynamical system is given by $\mathbf{x}^{*}=-A^{-1} \alpha$,  where the $N \times N$ matrix $A$ is given by $A = (A_{ij})$. This equilibrium is  globally asymptotically stable if and only if  $A$ is negative definite \cite{Grilli_NatComm2017}. Therefore, we set $A_{ii}=-c$, where $c$ is a constant that makes matrix $A$ negative definite. For simplicity, we set $c=\lambda_{\rm max}+1$, where $\lambda_{\rm max}$ is the largest eigenvalue of the adjacency matrix of the network. 

\begin{figure}
\includegraphics[height=!,width=0.9\textwidth]{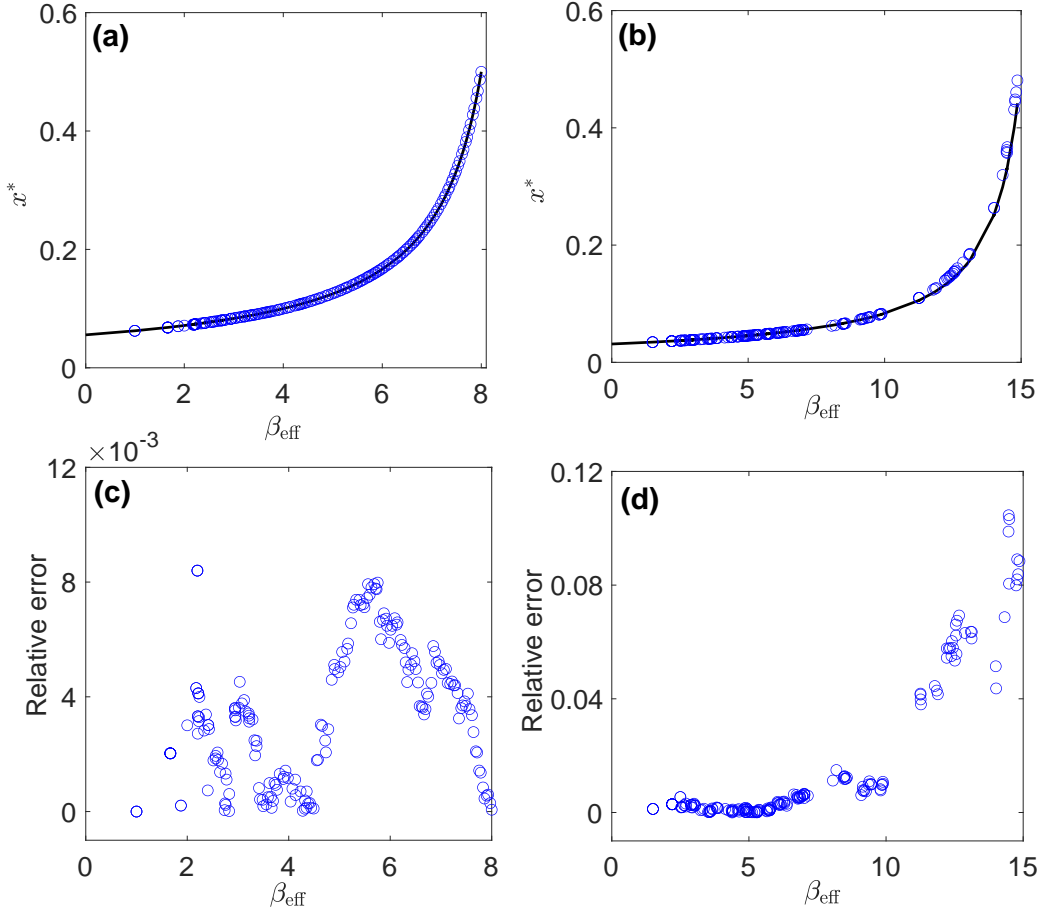}
\caption{{\bf GBB  reduction for the generalized Lotka-Volterra model.} (a) Bifurcation diagram for the regular random graph. (b) Bifurcation diagram  for the scale-free network. (c) Relative error for the regular random graph.  (d) Relative error for the scale-free network. The circles represent the numerically obtained equilibria for the initial condition $x_i=10$ for all $i$. The solid curves in (a) and (b) represent the stable equilibria of the GBB  reduction. }
\label{Figure51}
\end{figure}

We rewrite Eq.~\eqref{GLVeq1} as 
\begin{eqnarray}
\frac{dx_i}{dt}=\alpha x_i -cx_i^2+\sum_{j=1; j\neq i}^N A_{ij}x_ix_j.
\label{GLVeq11}
\end{eqnarray}
The GBB  reduction for Eq.~\eqref{GLVeq11} is given by 
\begin{eqnarray}
\frac{dx}{dt}=\alpha x + (\beta-c) x^2.
\label{GLVeq1D}
\end{eqnarray}
In Figs.~\ref{Figure51}(a) and \ref{Figure51}(b), we plot the bifurcation diagrams for the regular random graph and scale-free network, respectively, for the dynamical system given by Eq.~\eqref{GLVeq11} and its GBB  reduction given by Eq.~\eqref{GLVeq1D}. The relative error of the GBB  reduction corresponding to Figs.~\ref{Figure51}(a) and \ref{Figure51}(b)  is plotted in Figs.~\ref{Figure51}(c) and \ref{Figure51}(d), respectively.
We find that the GBB  reduction is fairly accurate across the examined range of $\beta_{\mathrm{eff}}$ for the regular random graph and that it is accurate when $\beta_{\mathrm{eff}}$ is sufficiently small in the case of the scale-free network.

\begin{figure}
\includegraphics[height=!,width=0.9\textwidth]{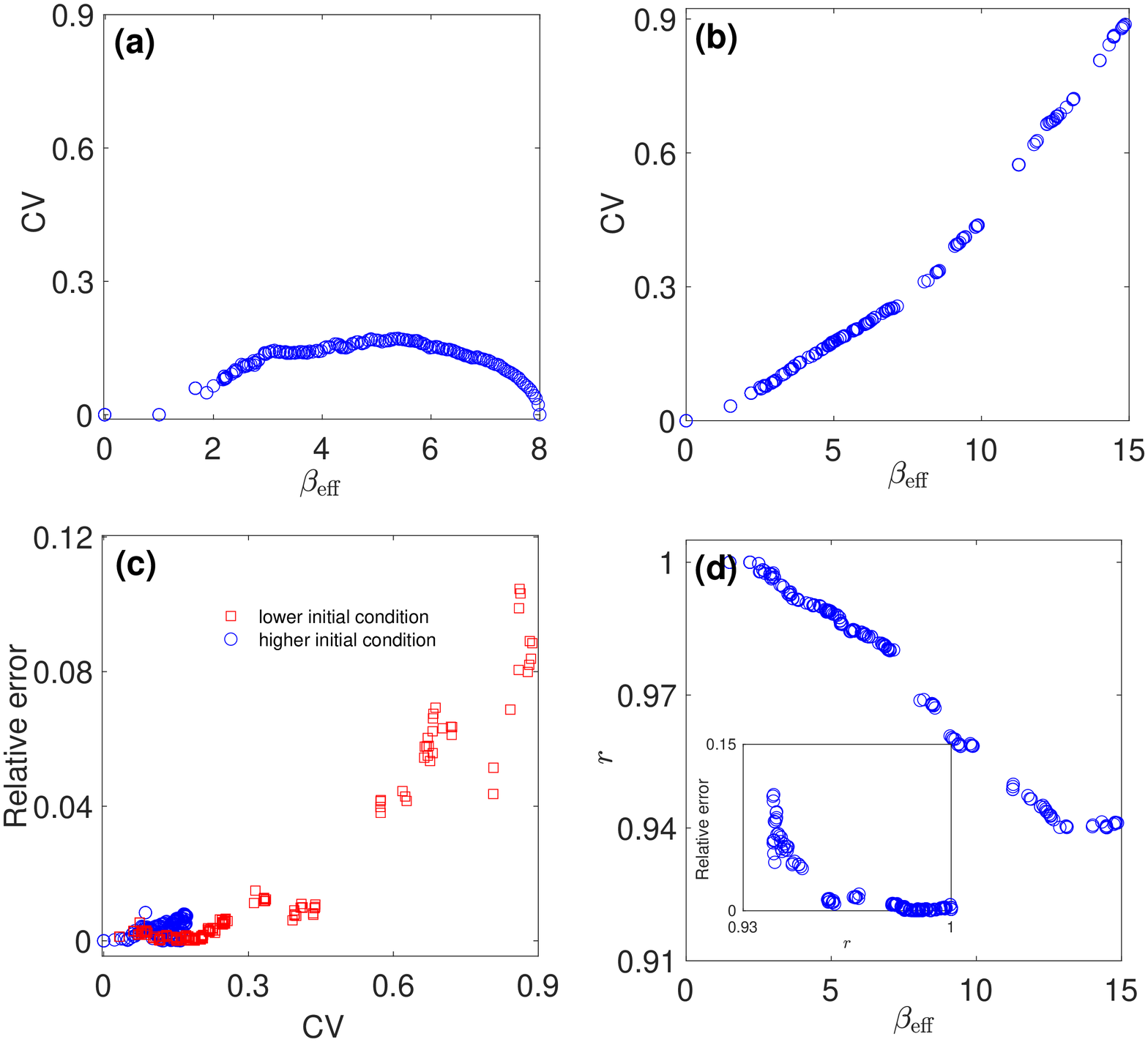}
\caption{ {\bf Exploring reasons for the accuracy of the GBB  reduction for the generalized Lotka-Volterra model.} 
 (a) CV of $\{x^*_1, \ldots, x^*_N\}$ for the regular random graph. (b) CV of $\{x^*_1, \ldots, x^*_N\}$ for the scale-free network. (c)  Relative error as a function of the CV for the two networks. 
(d) Pearson correlation coefficient between $x^*_i$ and $k_i$, i.e., $r$, as a function of $\beta_{\rm eff}$ for the scale-free network. The inset shows the relationship between the relative error and $r$.} 
\label{Figure52}
\end{figure}

For this dynamical system, we obtain $F(x_i)=\alpha x_i -cx_i^2$ and $G(x_i,x_j)=x_ix_j$.
We obtain
\begin{eqnarray}
R^{(1)}&=&\frac{F\left(\mathcal{L}\left(\mathbf{x}\right)\right)}{\mathcal{L}\left(F\left(\mathbf{x}\right)\right)}\nonumber\\
&=&\frac{ \frac{\langle s^{\rm{out}}{x}\rangle}{\langle s^{\mathrm{out}}\rangle}\left( \alpha-c\frac{\langle s^{\mathrm{out}}{x}\rangle}{\langle s^{\mathrm{out}}\rangle}\right)    }{\frac{\langle s^{\mathrm{out}}{x}\left(\alpha -c{x}\right)\rangle }{\langle s^{\mathrm{out}}\rangle} }\nonumber\\
&=&\frac{ {\alpha \langle s^{\mathrm{out}}{x}\rangle} \langle s^{\mathrm{out}}\rangle- c  {\langle s^{\mathrm{out}}{x}\rangle}^2 } {\alpha\langle s^{\mathrm{out}}{x}\rangle\langle s^{\mathrm{out}}\rangle - c\langle{s}^{\mathrm{out}}{x}^2\rangle \langle s^{\mathrm{out}}\rangle } .
\label{GLVapf1}
\end{eqnarray}
Equation ~\eqref{aprox2} holds with equality (i.e., $R^{(2)} = 1$) because 
\begin{eqnarray}
\mathcal{L}\left(G\left(x_i,\mathbf{x}\right)\right)&=&\mathcal{L}\left(x_i\mathbf{x} \right)  \nonumber \\
&=&x_i\mathcal{L}\left(\mathbf{x} \right)  \nonumber \\
&=&G\left(x_i,\mathcal{L}\left(\mathbf{x}\right)\right).
\end{eqnarray}
For the third approximation (see Eq.~\eqref{aprox3}), we obtain
\begin{eqnarray}
R^{(3)}&=&\frac{\mathcal{L}\left(\mathbf{s}^{\mathrm{in}}\right)G\left(\mathcal{L}\left(\mathbf{x}\right),\mathcal{L}\left(\mathbf{x}\right)\right)}{\mathcal{L}\left(\mathbf{s}^{\mathrm{in}}\circ  G\left(\mathbf{x},\mathcal{L}\left(\mathbf{x}\right)\right)\right)} \nonumber \\
&=&\frac{ \frac{\langle {s}^{\mathrm{in}}{s}^{\mathrm{out}}\rangle}{\langle {s}^{\mathrm{out}}\rangle}  \left(\frac{\langle {s}^{\mathrm{out}}{x}\rangle}{\langle {s}^{\mathrm{out}}\rangle}\right)^2 } {\frac{\langle {s}^{\mathrm{in}} {s}^{\mathrm{out}} {x}\rangle\langle {s}^{\mathrm{out}}{x}\rangle}{\langle {s}^{\mathrm{out}}\rangle^2}  } \nonumber\\
&=&\frac{ \frac {\langle {s}^{\mathrm{out}}{x}\rangle} {\langle {s}^{\mathrm{out}}\rangle}} {\frac {\langle {s}^{\mathrm{in}} {s}^{\mathrm{out}} {x}\rangle}{\langle {s}^{\mathrm{in}}{s}^{\mathrm{out}}\rangle} }\,.
\label{GLVap1}
\end{eqnarray}
Equations \eqref{GLVapf1} and \eqref{GLVap1} imply that the GBB  reduction is exact for the uncorrelated networks if $x_i$ is the same for all $i$. 
%
%
In Figs.~\ref{Figure52}(a) 
and \ref{Figure52}(b), we plot the CV of $\{ x^*_1, \ldots, x^*_N \} $ as a function of  $\beta_{\mathrm{eff}}$ for the regular random graph and scale-free network, respectively. The CV is not monotonic in the case of the regular random graph. In contrast, the CV increases with the increase in $\beta_{\mathrm{eff}}$ for the scale-free network. We plot the relative error as a function of CV for the two networks in Fig.~\ref{Figure52}(c). We find that the relationship between the relative error and CV is approximately monotone and quantitatively similar between the two networks. 
The pattern of the correlation between
$x_i$ and $k_i$ for the scale-free network shown in Fig.~\ref{Figure52}(d) is similar to 
that shown in Fig.~ \ref{Figure22}(d) in that the correlation is large regardless of $\beta_{\rm eff}$ and that the relative error decreases as the correlation increases. Therefore, the correlation between $x_i$ and $k_i$ does not explain the accuracy of the GBB  reduction. 
We conclude that, as in the case of the double-well system and the SIS model, the CV of $\{x^*_1,\dots,x^*_N\}$ mainly determines the accuracy of the GBB  reduction.

\subsection{Mutualistic dynamics}\label{M_sys_sec}

Finally, we consider the  mutualistic interaction dynamics among different species in ecological networks given by
\begin{eqnarray}
\frac{dx_i}{dt}=B_i&+&x_i\left(1-\frac{x_i}{K_i}\right)\left(\frac{x_i}{C_i}-1\right)\nonumber \\ &+&\sum_{j=1}^N A_{ij}\frac{x_ix_j}{D_i+E_ix_i+H_jx_j},
\label{M_system1e}
\end{eqnarray}
where $x_i$ represents the abundance of the species $i$, and $B_i,~ C_i,~ D_i,~ E_i,~ H_i,$ and $K_i$ with $i=1, \ldots, N$  are constants \cite{Gao_Nature2016}. The first term  on the right-hand side of Eq.~\eqref{M_system1e}, i.e., $B_i$, represents the migration rate of the species $i$ from  outside the ecosystem. The second term stands for the logistic growth with the carrying capacity $K_i$, and $C_i$ represents the Allee constant. 
The third term represents the mutualistic interaction term, i.e., the contribution of $x_j$ to $x_i$. This term remains bounded because the logistic growth (i.e., the second term) does not allow $x_i$ to excessively grow beyond $K_i$.

\begin{figure}
\centering
\includegraphics[height=!,width=0.9\textwidth]{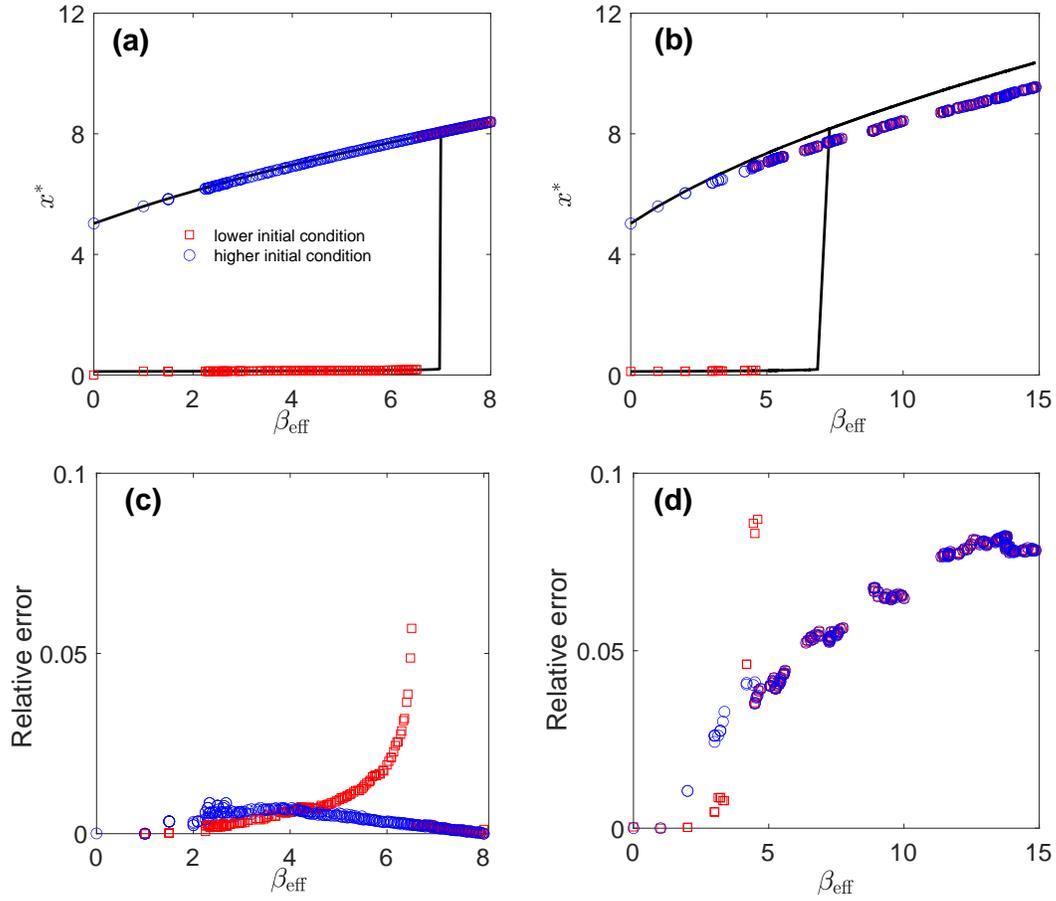}
\caption{{\bf GBB  reduction for the mutualistic system.} (a) Bifurcation diagram for the regular random graph. (b) Bifurcation diagram of the  scale-free network.  (c) Relative error for the regular random graph.  (d) Relative error for the scale-free network.  The squares and circles correspond to the numerically obtained equilibria when the initial condition is $x_i=0.01$ and $x_i=10$, respectively, with $i=1,\dots,N$. The solid lines in (a) and (b) represent the stable equilibria of the GBB  reduction given by  Eq.~\eqref{M_system1e}.} 
\label{M_sysFigure11}
\end{figure}

Note that the present model  is a nonlinear-interaction variant  of the double-well system considered in Sec. \ref{double_well}. 
We follow Ref.~\cite{Gao_Nature2016} to set $B_i=B=0.1$, $C_i=C=1$, $K_i=K=5$, $D_i=D=5$, $E_i=E=0.9$, and 
$H_i=H=0.1$ for all $i$. 
The GBB  reduction is given by 
\begin{eqnarray}
\frac{dx}{dt}=B+x\left(1-\frac{x}{K}\right)\left(\frac{x}{C}-1\right) + \frac{\beta x^2}{D+Ex+Hx}\,.\nonumber \\
\label{M_system1D}
\end{eqnarray}
 
The bifurcation diagrams for the regular random graph and the scale-free network are shown in Figs.~\ref{M_sysFigure11}(a) and \ref{M_sysFigure11}(b), respectively, for both the original dynamical system given by Eq.~\eqref{M_system1e}  and its GBB  reduction given by Eq.~\eqref{M_system1D}.
The GBB  reduction is inaccurate at estimating the bifurcation point for both networks, in particular for the scale-free network (see Fig.~\ref{M_sysFigure11}(b)).
The relative error of the GBB  reduction corresponding to Figs.~\ref{M_sysFigure11}(a) and \ref{M_sysFigure11}(b) is shown in Figs.~\ref{M_sysFigure11}(c) and \ref{M_sysFigure11}(d), respectively. We find that the relative error is small except near the bifurcation point for both networks.

\begin{figure}
\centering
\includegraphics[height=!,width=0.9\textwidth]{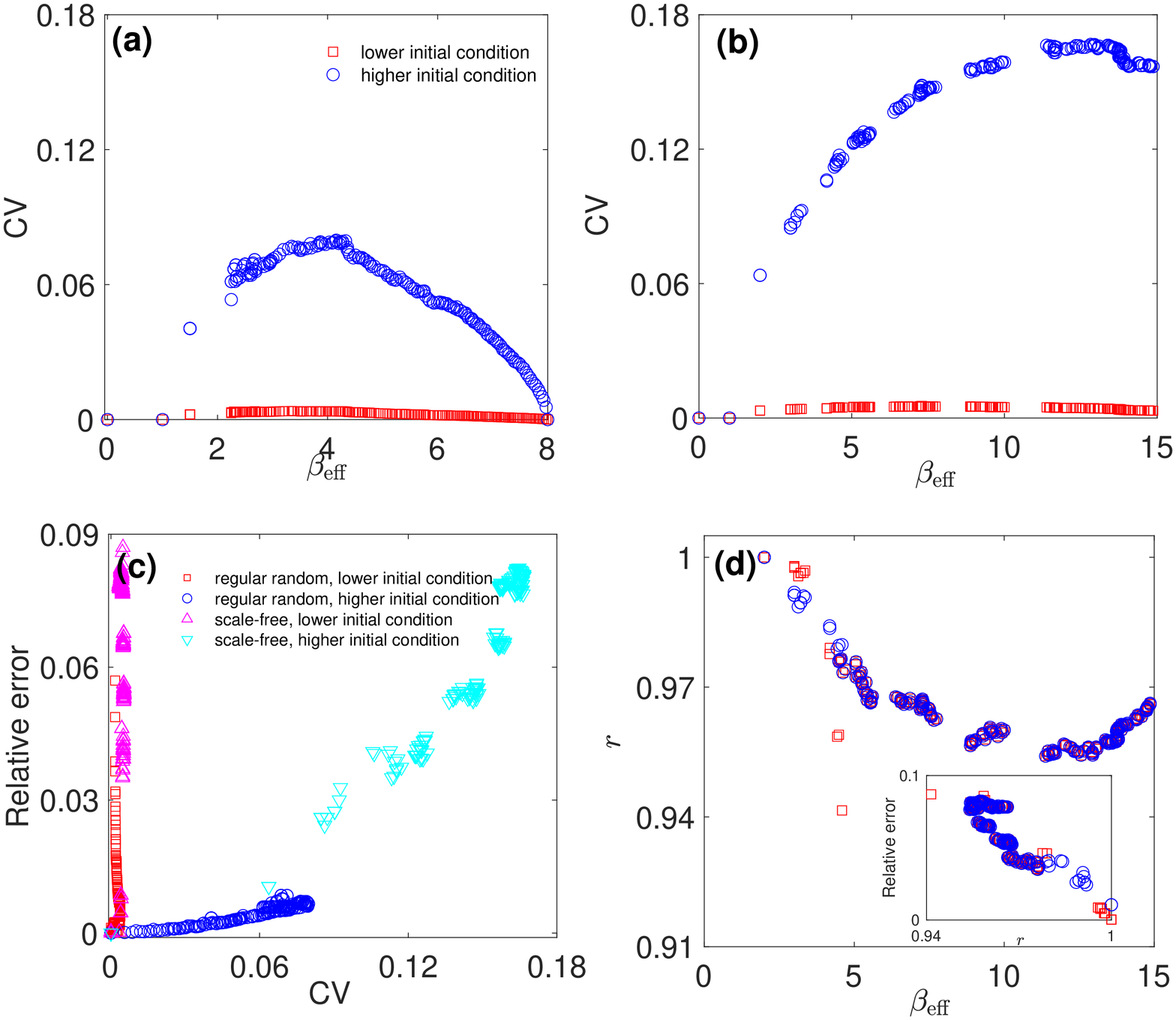}
\caption{{\bf Exploring reasons for the accuracy of the GBB  reduction for the mutualistic system. }(a) CV of $\{x^*_1, \ldots, x^*_N\}$ for the regular random graph. (b) CV of $\{x^*_1, \ldots, x^*_N\}$ for the scale-free network. (c) Relative error as a function of the CV for  the two networks. (d) Pearson correlation coefficient between $x^*_i$ and $k_i$, i.e., $r$, for the scale-free network as a function of $\beta_{\mathrm{eff}}$. The inset shows the relationship between the relative error and $r$.} 
\label{M_sysFigure12}
\end{figure}

Equation \eqref{M_system1e} implies
 \begin{eqnarray}
F(x_i)=B_i+x_i\left(1-\frac{x_i}{K_i}\right)\left(\frac{x_i}{C_i}-1\right)
\end{eqnarray}
and 
 \begin{eqnarray}
G\left(x_i,x_j\right)=\frac{x_ix_j}{D_i+E_ix_i+H_jx_j}\,.
\end{eqnarray}
Therefore, we obtain 
\begin{widetext}
\begin{eqnarray}
R^{(1)}&=&\frac{F\left(\mathcal{L}\left(\mathbf{x}\right)\right)}{\mathcal{L}\left(F\left(\mathbf{x}\right)\right)}\nonumber \\
&=&\frac{B+\frac{\langle {s}^{\mathrm{out}}{x}\rangle}{\langle {s}^{\mathrm{out}}\rangle}\left(1-\frac{\langle {s}^{\mathrm{out}}{x}\rangle}{K\langle {s}^{\mathrm{out}}\rangle}\right)\left(\frac{\langle {s}^{\mathrm{out}}{x}\rangle}{C\langle {s}^{\mathrm{out}}\rangle}-1\right)}{\frac{\left\langle {s}^{\mathrm{out}} \left(B+{x}\left(1-\frac{{x}}{K}\right)\left(\frac{{x}}{C}-1\right)\right)\right\rangle}{\langle {s}^{\mathrm{out}}\rangle}}\nonumber\\
&=&\frac{BKC \langle {s}^{\mathrm{out}}\rangle^3 -KC\langle {s}^{\mathrm{out}}\rangle^2 \langle {s}^{\mathrm{out}}{x}\rangle +\left(K+C\right) \langle {s}^{\mathrm{out}}\rangle \langle {s}^{\mathrm{out}}{x}\rangle^2-\langle {s}^{\mathrm{out}}{x}\rangle^3}{BKC \langle {s}^{\mathrm{out}}\rangle^3- KC \langle {s}^{\mathrm{out}} \rangle^2 \langle {s}^{\mathrm{out}} {x}\rangle +  \left(K+C\right) \langle {s}^{\mathrm{out}}\rangle^2 \langle {s}^{\mathrm{out}}{x}^2\rangle  -\langle {s}^{\mathrm{out}}\rangle^2 \langle {s}^{\mathrm{out}}{x}^3\rangle }\,,
\label{M_system1}\\
R^{(2)}&=&\frac{G\left(x_i,\mathcal{L}\left(\mathbf{x}\right)\right)}{\mathcal{L}(G\left(x_i,\mathbf{x}\right)}\nonumber\\
&=&\frac{\langle {s}^{\mathrm{out}}{x}\rangle}{\left(D+Ex_i+H\frac{\langle {s}^{\mathrm{out}}{x}\rangle}{\langle {s}^{\mathrm{out}}\rangle}\right)\left( \left\langle {s}^{\mathrm{out}}\frac{{x}}{D+Ex_i+H{x}}\right\rangle\right)}  \label{M_system2}\,,
\end{eqnarray}
and
\begin{eqnarray}
  R^{(3)}&=&\frac{\mathcal{L}(\mathbf{s}^{\mathrm{in}})G(\mathcal{L}(\mathbf{x}),\mathcal{L}(\mathbf{x}))} {\mathcal{L}(\mathbf{s}^{\mathrm{in}}\circ  G(\mathbf{x},\mathcal{L}(\mathbf{x})))}\nonumber\\
  &=&\frac{\frac{\langle {s}^{\mathrm{in}}{s}^{\mathrm{out}}\rangle\frac{\langle {s}^{\mathrm{out}}{x}\rangle^2}{\langle {s}^{\mathrm{out}}\rangle^2}}{\langle {s}^{\mathrm{out}}\rangle\left(D+E\frac{\langle {s}^{\mathrm{out}}{x}\rangle}{\langle {s}^{\mathrm{out}}\rangle}+H\frac{\langle {s}^{\mathrm{out}}{x}\rangle}{\langle {s}^{\mathrm{out}}\rangle}\right)}
}{\frac{\left\langle \frac{{s}^{\mathrm{in}}{s}^{\mathrm{out}}{x}\frac{\langle {s}^{\mathrm{out}}{x}\rangle}{\langle {s}^{\mathrm{out}}\rangle}}{D+E{x}+H\frac{\langle {s}^{\mathrm{out}}{x}\rangle}{\langle {s}^{\mathrm{out}}\rangle}}\right\rangle}{\langle {s}^{\mathrm{out}}\rangle}} \nonumber \\
&=&{\frac{\langle {s}^{\mathrm{in}}{s}^{\mathrm{out}}\rangle{\langle {s}^{\mathrm{out}}{x}\rangle}}{D\langle {s}^{\mathrm{out}}\rangle+E\langle {s}^{\mathrm{out}}{x}\rangle+H\langle {s}^{\mathrm{out}}{x}\rangle}
}\frac{1}{\left\langle \frac{{s}^{\mathrm{in}}{s}^{\mathrm{out}}{x}}{D+E{x}+H\frac{\langle {s}^{\mathrm{out}}{x}\rangle}{\langle {s}^{\mathrm{out}}\rangle}}\right\rangle}\,.
 \label{M_system3}
\end{eqnarray}
\end{widetext}
In the case of the mutualistic dynamics, none of the three  approximations holds with equality without a further condition.  
Equations \eqref{M_system1}, \eqref{M_system2}, and \eqref{M_system3} imply that all the approximations are exact for uncorrelated networks when all $x_i$'s are the same.
In Figs.~ \ref{M_sysFigure12}(a) and  \ref{M_sysFigure12}(b), we plot the CV of $\{x^*_1,\dots,x^*_N\}$ as a function of $\beta_{\rm eff}$ for the regular random graph and scale-free network, respectively. The  squares and the circles correspond to the lower and higher  equilibria, respectively. The CV is small (i.e., less than 0.18) for both networks. We plot the relative error against the CV  for the two networks in Fig.~ \ref{M_sysFigure12}(c).  The figure shows that the magnitude of the relative error is strongly correlated with the CV of  $\{x^*_1,\dots,x^*_N\}$. This result is consistent with the observation that the magnitude of the relative error is smaller for the regular random graph than the scale-free network (compare Figs.~\ref{M_sysFigure11}(c) and  \ref{M_sysFigure11}(d)).
We also confirm that the Pearson correlation coefficient between $x^*_i$ and $k_i$, shown in Fig.~\ref{M_sysFigure12}(d), is large irrespective of the $\beta_{\rm eff}$ value and that the relative error is negatively rather than positively related with the correlation coefficient. These results are qualitatively the same as those shown in Figs.~\ref{Figure22}(d) and \ref{Figure52}(d).
Therefore, we conclude that the correlation between $x_i$ and $k_i$ does not explain the accuracy of the GBB  reduction
and that the CV of $\{x^*_1,\dots,x^*_N\}$ mainly determines the accuracy of the GBB  reduction for the  mutualistic dynamics given by Eq.~\eqref{M_system1e}.

\section{Discussion}

In this paper, we examined when the GBB  reduction of $N$-dimensional dynamical systems on networks \cite{Gao_Nature2016} is accurate.
The high accuracy of the approximation clearly hinges upon the assumption of uncorrelated networks, similar to the case of heterogeneous meanfield approximations for percolation and various dynamics on networks.
We showed that, apart from that, a main determinant of the accuracy of the GBB  reduction is the CV of the relevant dynamical variable, which depends on the considered dynamical system, in the equilibrium. In four out of the five dynamical systems, i.e., 
the double-well system, SIS model, GLV model, and mutualistic system,
the CV of $\{x^*_1, \dots, x^*_n\}$ was positively correlated with the approximation error. For the gene regulatory system, the CV of $\left\{(x^*_1)^2/(1+(x^*_1)^2), \ldots, (x^*_N)^2/(1+(x^*_N)^2) \right\}$ was a better predictor of the accuracy of the approximation than the CV of $\{x^*_1,\dots,x^*_N\}$. We suggest that, given the dynamical system on networks, one should look at the dependence of the interaction term on the state variables to determine the node-dependent quantity of which the CV is to be considered.
We also found that the accuracy of the GBB  reduction was generally higher for the regular random graph than the scale-free network for all five dynamical systems. This result is consistent with the dependence of the accuracy of the GBB  reduction on the CV because the state variable in the equilibrium tends to be more homogeneous for the regular random graph than for the scale-free network.
Moreover, we also showed that the high correlation between the state variable and the node's degree, which was present in all five dynamical systems, did not affect the accuracy of the GBB  reduction.  Here we summarize the result in  Table \ref{table1}. We summarize here the dynamical properties which decides the accuracy of the GBB reduction. We also show which approximation/approximations are exact and which are responsible for the accuracy of the GBB reduction. 
  \begin{center}
\begin{table}[h!]
    \label{tab:table1}
    \begin{tabular}{|m{2cm}| m{2cm} | m{1.2cm}|m{1.2cm}|m{1.2cm}|}
     \hline
      Dynamical system  & Accuracy of GBB depends on  & $R^{(1)}=1$& $R^{(2)}=1$& $R^{(3)}=1$\\
      \hline
      Double-well system & CV($x$) & No & Yes& Yes\\
       \hline
       SIS model & CV($x$) & Yes & Yes& No\\
        \hline
       Gene regulatory system & CV($\frac{x}{1+x^2}$) & Yes & No& Yes\\
        \hline
       GLV model & CV($x$) & No & Yes& No\\
        \hline
       Mutualistic system & CV($x$) & No & No& No\\
        \hline
    \end{tabular}
 \caption{\label{table1} The dynamical property which is responsible the accuracy of the GBB reduction and the approximations on which the  the GBB reduction depends for different dynamical systems.}
\end{table}
  \end{center}

Laurence and coauthors proposed a generalized dimension reduction method for dynamical systems on networks given in the form of Eq.~\eqref{multi_eq1}, with which the effective state is a general linear weighted sum of the nodes' states, i.e., $x = \sum_{i=1}^N a_i x_i$ for constants $a_i$~\cite{Laurence_PRX2019}. The GBB  reduction is a special case of this generalized reduction. 
The generalized reduction is optimal when 
$(a_1, \ldots, a_N)$ is the dominant eigenvector of the adjacency matrix of the network~\cite{Laurence_PRX2019}.
We refer to the one-dimensional reduction $x = \sum_{i=1}^N a_i x_i$ with this particular $(a_1, \ldots, a_N)$ as the reduction using spectral method. The GBB  reduction coincides with the reduction using spectral method for uncorrelated networks.
We have numerically examined the accuracy of the reduction using spectral method for the five dynamical systems in the Appendix~\ref{sec:Laurence}. We find that the relative error obtained with the reduction using spectral method depends on the dynamical system. The relative error is smaller for the reduction using spectral method than the GBB  reduction for the double-well, gene regulatory and mutualistic systems, whereas the opposite is the case for the SIS and GLV models. Another observation is that the reduction using spectral method does not accurately locate the bifurcation point for some dynamical systems. It does locate the bifurcation point at a higher accuracy than the GBB  reduction in the case of the mutualistic dynamics. However, the reduction using spectral method does not accurately locate the bifurcation point in the case of the double-well system, similar to the GBB  reduction.  For SIS and gene regulatory systems both GBB  and reduction using spectral methods accurately locate the bifurcation point. (The GLV model does not show a bifurcation.) We also remark that the two reduction methods are similar in that they are accurate when the CV of $\{x^*_1, \dots, x^*_n\}$ (or of $\{(x^*_1)^2/(1+(x^*_1)^2), \ldots,
(x^*_N)^2/(1+(x^*_N)^2) \}$ in the case of the gene regulatory dynamics) is small. To conclude, we cannot say that the reduction using spectral method is uniformly better than the GBB  reduction. Further developing low-dimensional reduction methods that
accurately predict the bifurcation point for wider classes of dynamical systems remains an open question.

The GBB  and reduction using spectral methods require that neither the intrinsic node's dynamics (i.e., function $F$) nor the coupling term (i.e., function $G$) depends on the node. In a recent study, Tu {\it et al.} proposed a dimension reduction technique that is applicable to cases in which $F$ or $G$ depends on nodes~\cite{Tu_IScience2021}. While this new approach accommodates realistic scenarios such as the case in which external input is injected to a subset of nodes \cite{Liu_Nature2011, Cramer_PlosOne2016},
the validity of this approach requires that the degree distribution of the network is not highly heterogeneous, the CV of the state variable is small, and the parameters of $F$ and $G$ are not too heterogeneous \cite{Tu_IScience2021}.
Developing low-dimensional reductions that remove any of these restrictions warrants future work.  
%

In the case of the GLV model, the negative definiteness of the interaction matrix, $A$, is a necessary and sufficient condition  for the stability of the nontrivial equilibrium \cite{Tu_PRE2017, Grilli_NatComm2017}. In Ref.~\cite{Tu_PRE2017}, the authors assumed that ($A_{ij}$, $A_{ji}$) with $i\neq j$ is chosen from a bivariate distribution with the mean of each variable equal to $\mu$, standard deviation of each variable equal to $\sigma$, and correlation coefficient between $A_{ij}$ and $A_{ji}$ equal to $\rho$. To carry out analytical calculations, the authors also assumed that $A_{ii}=-d \leq -d_{\text{c}}$ for all $i\in \{1, \ldots, N\}$, where $d_{\text{c}} = \sigma\sqrt{2N(1+\rho)}-\mu$ when $\mu\leq0$ and $d_{\text{c}}=(N-1)\mu$ when $\mu>0$. This choice renders $A$ negative definite. Then, they assumed that the $A_{ij}G(x_i, x_j)$ terms for all $i$ and $j$ enter the coupling function $G$ in the GBB  reduction. In contrast, we proposed that, in the GBB  reduction, the self-interaction term, $A_{ii}G(x_i, x_i)$, should enter function $F$, which describes the inherent dynamics of the node, rather than $G$. Therefore, we needed to assume that $A_{ii}$ is independent of $i$, which is a condition for the validity of the GBB  reduction. Although this assumption coincides with  $A_{ii} = -d$ assumed in Ref.~\cite{Tu_PRE2017}, the reason for assuming this is different between their work and ours. It should be noted that this limitation has been removed in their subsequent work, where the authors developed a reduction method that allows node-dependent $F$ and $G$~\cite{Tu_IScience2021}.

The GBB  reduction is not applicable to the case of diffusive coupling. This is because the right-hand side of Eq.~\eqref{aprox3} vanishes for diffusive coupling, whereas the left-hand side
does not in general. However, there is a huge demand of understanding dynamical systems in which the nodes are diffusively coupled in networks. For example, the Kuramoto model of coupled oscillators often assumes diffusive coupling between oscillators, and its network versions have been broadly investigated \cite{Arenas_PhysReport2008, Rodrigues_PhysReport2016}.
While dimension reduction techniques for coupled oscillator systems on networks are available \cite{Ichinomiya_PRE2004, Kawamura_PRL2008, Kori_PRE2009, Rodrigues_PhysReport2016, Ji_SciRep2014}, they typically reduce
the original dynamics to lower yet high-dimensional dynamical systems; for example, the dimension of the reduced dynamical system is equal to the number of distinct values of the node's degree. The development of dimension reduction methods that are applicable to more general dynamics on networks with diffusive coupling and those that realize lower-dimensional representations of the original dynamics in a theoretically principled manner is also an open challenge.

\section*{Acknowledgments}
Authors would like to thank Vincent Thibeault and Baruch Barzel for fruitful discussions.
H.K. acknowledges support from JSPS KAKENHI (under Grant No. 21K12056).
N.M. acknowledges support from AFOSR European Office
(under Grant No. FA9550-19-1-7024), the Sumitomo Foundation, and the 
Japan Science and Technology Agency (JST) Moonshot R\&D (under Grant No. JPMJMS2021).


\appendix

\section{\label{sec:Laurence}Laurence's one-dimensional reduction}

The method proposed by Laurence and colleagues~\cite{Laurence_PRX2019} reduces the $N$-dimensional dynamical system considered in Eq.~\eqref{multi_eq1} to a one-dimensional dynamical system given by
 \begin{eqnarray}
   \dot{x}=F(x)+\alpha_{\rm ev}G(\beta_{\rm sp} x,x),
   \label{1Dprx}
 \end{eqnarray}
where $\alpha_{\rm ev}$ is the dominant eigenvalue of the adjacency matrix of the network in terms of the modulus. We assume that the  dominant eigenvalue  is reasonably larger than the second largest eigenvalue in terms of modulus.
One assumes that
\begin{eqnarray}
   x=\sum_{i}^N b_ix_i=\mathbf{b}^{\top} \mathbf{x},
\end{eqnarray}
where $\mathbf{b}$ is the eigenvector corresponding to $\alpha_{\rm ev}$ and normalized as $\sum_{i=1}^N b_i=\mathbf{1^\top b}=1$. The parameter $\beta_{\rm sp}$ only depends on the network structure and is given by 
\begin{eqnarray}
   \beta_{\rm sp}=\frac{\mathbf{b}^\top \mathbf{K}\mathbf{b}}{\alpha_{\rm ev}\mathbf{b}^\top\mathbf{b}},
\end{eqnarray}
where $\mathbf{K} = (K_{ij})$ is the $N\times N$ diagonal matrix with $K_{ii}=k_{i}^{\rm in}$. 
In the case of uncorrelated random networks, we obtain 
\begin{eqnarray}
   \alpha_{\rm ev}\approx \frac{\sum_{i=1}^N k_i^{\rm out}k_i^{\rm in}}{\sum_{i=1}^N k_i^{\rm out}}=\beta_{\rm eff},
\end{eqnarray}
and $\beta_{\rm sp}\approx 1$, with which Eq.~\eqref{1Dprx} coincides with the GBB  reduction~\cite{Laurence_PRX2019}.

\begin{figure}
\centering
\includegraphics[height=!,width=0.9\textwidth]{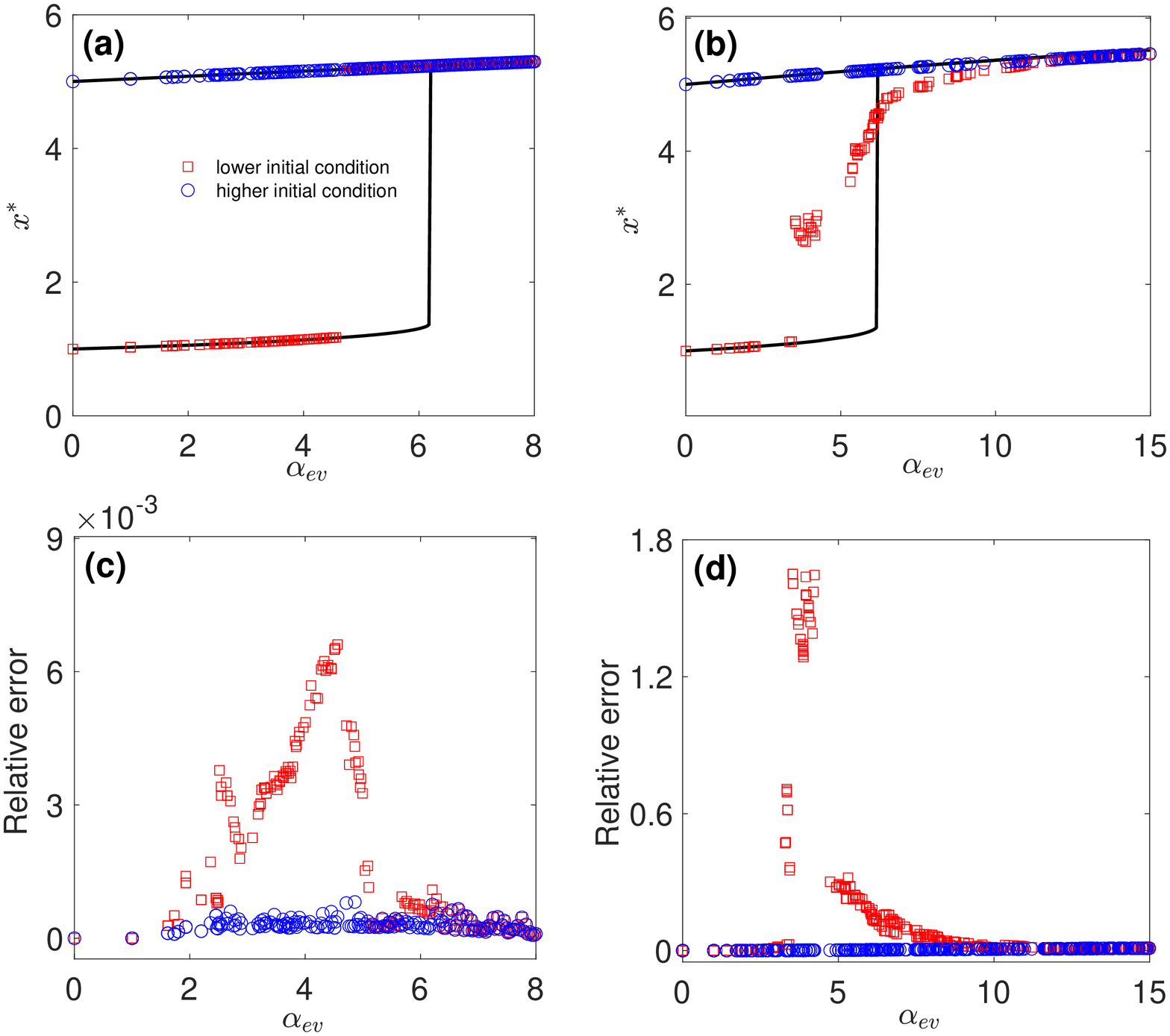}
\caption{{\bf  Reduction using spectral method for the double-well system.} (a) Bifurcation diagram for the regular random graph. (c) Bifurcation diagram for the scale-free network. (c) Relative error for the regular random graph. (d) Relative error for the scale-free network. The squares and circles represent the numerically obtained equilibria when the initial condition is $x_i=0.01$ or $x_i=10$, respectively, for all $i$. At each $\alpha_{\mathrm{ev}}$ value, we started the simulation of the original dynamical system from each of these two initial conditions and obtained the equilibria. The solid lines in (a) and (b) represent the stable equilibria of the one-dimensional reduction given by Eq.~\eqref{1Dprx} for the double-well system (i.e., Eq.~\eqref{double_well_system1}).
} 
\label{double_well_fig1prx}
\end{figure}

\begin{figure}
\includegraphics[height=!,width=0.9\textwidth]{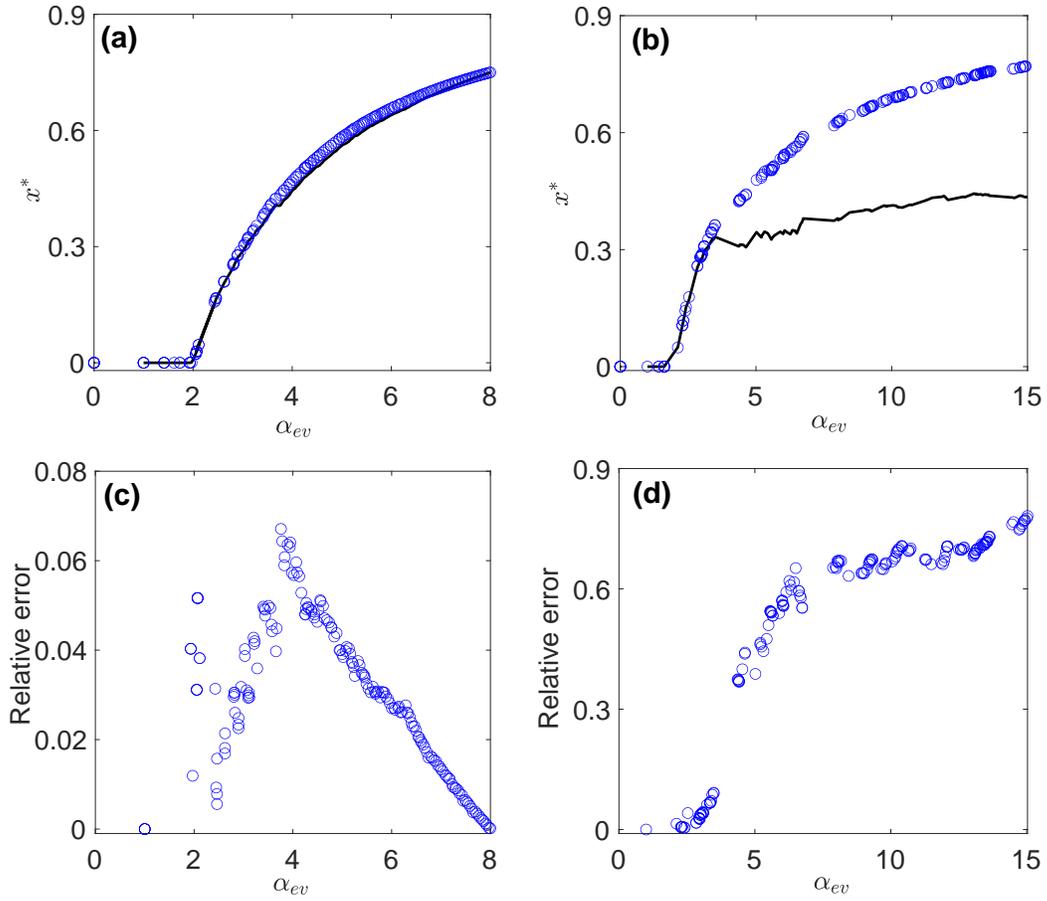}
\caption{{\bf Reduction using spectral method for the SIS model. }(a) Bifurcation diagram  for the regular random graph. (b) Bifurcation diagram  for the scale-free network. (c) Relative error  for the regular random graph.  (d) Relative error  for the scale-free network. The circles represent the numerically obtained equilibria  with the initial condition  $x_i= 10$ for all $i$. The solid curves in (a) and (b) represent the stable equilibria of the one-dimensional reduction given by Eq.~\eqref{1Dprx} for the SIS model (i.e., Eq.~\eqref{SISeq1}). } 
\label{SISFigure1prx}
\end{figure}

\begin{figure}
\includegraphics[height=!,width=0.9\textwidth]{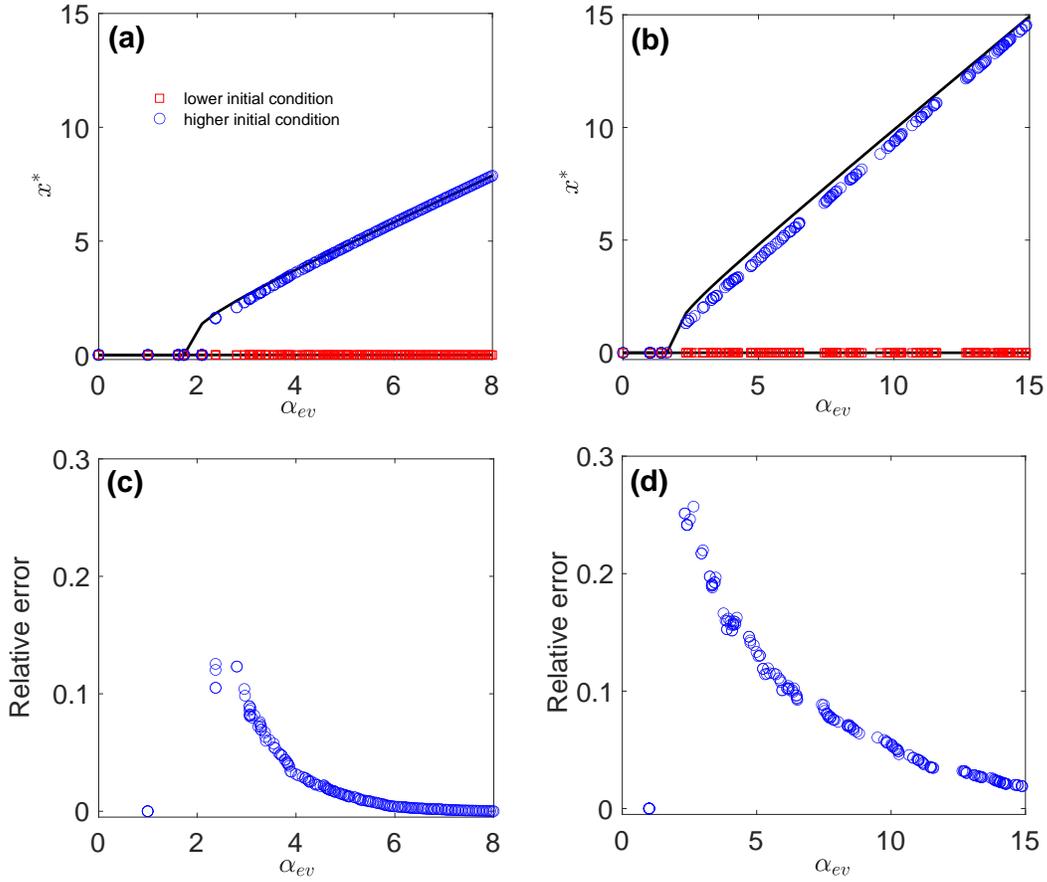}
\caption{{\bf  Reduction using spectral method for the gene regulatory system.} (a) Bifurcation diagram  for the regular random graph. (b) Bifurcation diagram for the scale-free network. (c) Relative error at the nontrivial equilibria for the regular random graph.   (d) Relative error at the nontrivial equilibria for the scale-free network. The solid curves in (a) and (b) represent the stable equilibria of the one-dimensional reduction given by Eq.~\eqref{1Dprx} for the gene regulatory system (i.e., Eq.~\eqref{R_system1}).}
\label{Figure21prx}
\end{figure}

\begin{figure}[h]
\includegraphics[height=!,width=0.9\textwidth]{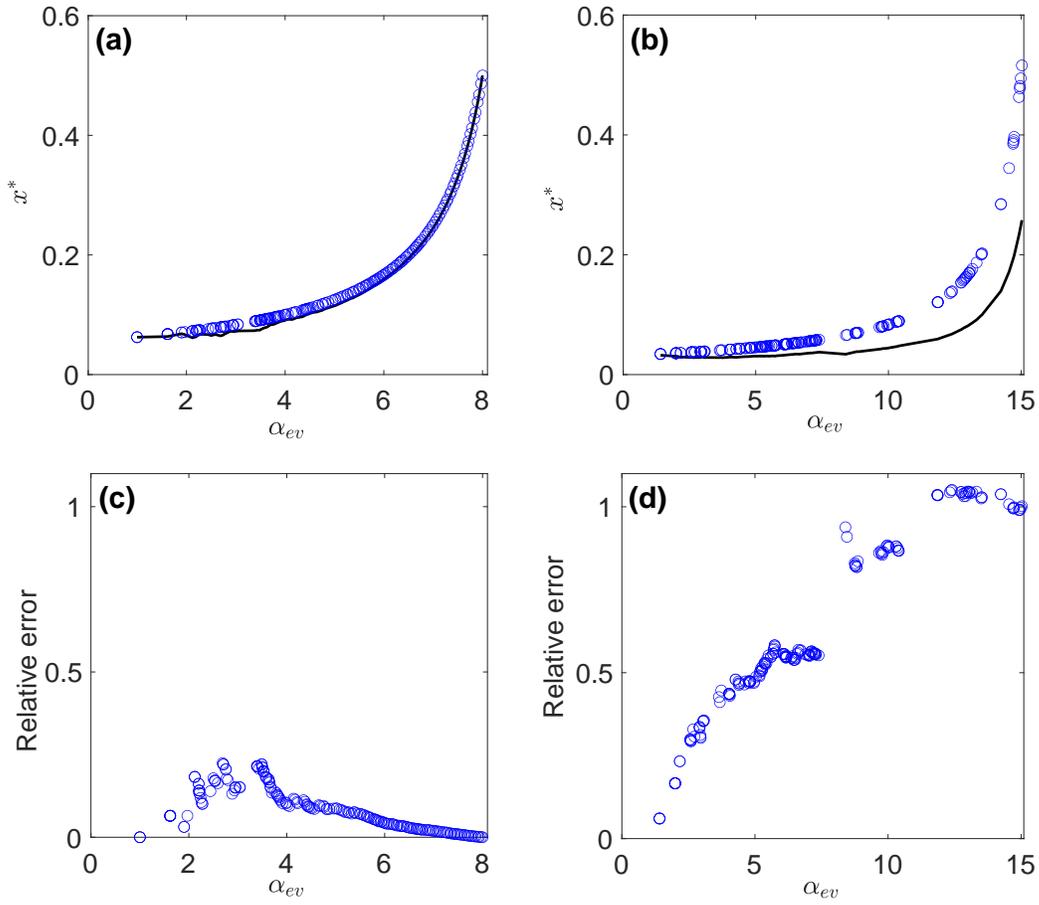}
\caption{{\bf  Reduction using spectral method for the generalized Lotka-Volterra model.} (a) Bifurcation diagram  for the regular random graph. (b) Bifurcation diagram  for the scale-free network. (c) Relative error for regular random graph.  (d) Relative error for scale-free network. The circles represent the numerically obtained equilibria for the initial condition $x_i=10$ for all $i$. The solid curves in (a) and (b) represent the stable equilibria of the one-dimensional reduction given by Eq.~\eqref{1Dprx} for the GLV model (i.e., Eq.~\eqref{GLVeq11}).}
\label{Figure51prx}
\end{figure}

\begin{figure}[h]
\centering
\includegraphics[height=!,width=0.9\textwidth]{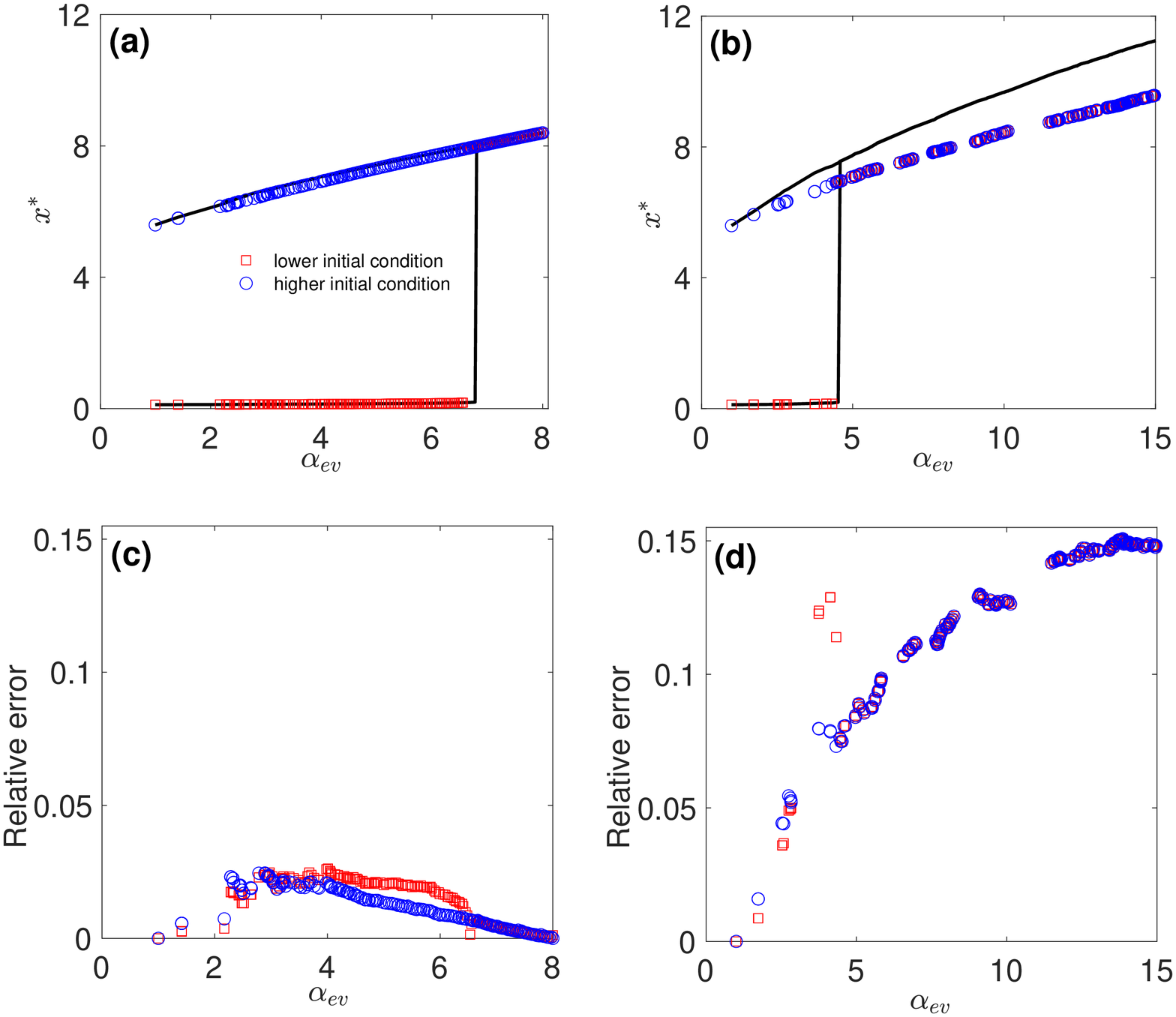}
\caption{{\bf  Reduction using spectral method for the mutualistic system.} (a) Bifurcation diagram for the regular random graph. (b) Bifurcation diagram of the scale-free network. (c) Relative error for the regular random graph.  (d) Relative error for the scale-free network.  The  squares and circles correspond to the numerically obtained equilibria when the initial condition is $x_i=0.01$ or $x_i=10$, respectively, for all $i$. The solid lines in (a) and (b) represent the stable equilibria of the one-dimensional reduction given by  Eq.~\eqref{1Dprx} for the mutualistic system (i.e., Eq.~\eqref{M_system1e}).} 
\label{M_sysFigure11prx}
\end{figure}

We test the accuracy of the reduction using spectral method on the five dynamical systems that we considered in the main text.
The bifurcation diagrams in terms of the effective state and the relative error, both as a function of $\alpha_{\rm ev}$, are shown in Figs.~\ref{double_well_fig1prx}--\ref{M_sysFigure11prx}. For the double-well system, gene regulatory system, and mutualistic system, the relative error with the reduction using spectral method is smaller than with the GBB  reduction, as shown in Figs.~\ref{double_well_fig1prx}, \ref{Figure21prx}, and \ref{M_sysFigure11prx}, respectively. In contrast, the opposite result holds true for the SIS and GLV models, as shown in Figs.~\ref{SISFigure1prx} and \ref{Figure51prx}, respectively.

\newpage
\clearpage

%


\end{document}